\documentclass[a4paper, 11pt]{article}

\usepackage{graphicx}
\usepackage{color}
\usepackage{lineno}
\usepackage{amssymb}
\usepackage{amssymb,latexsym,amsmath, graphicx, amsfonts, caption, color, verbatim, booktabs}
\usepackage{bbm}
\usepackage[bbgreekl]{mathbbol}
\usepackage{listings}
\usepackage{subcaption}
\usepackage{hyperref}

\RequirePackage[OT1]{fontenc}
\RequirePackage[numbers, compress]{natbib}
\usepackage{subcaption}
\numberwithin{equation}{section}

\title{
Conditional power, predictive power and probability of success in clinical trials with continuous, binary and time-to-event endpoints}

\author{Madan G Kundu$^1$\thanks{
        \textit{Corresponding author: Madan G. Kundu, {mkundu@dsi.com}}}
        \thanks{This article reflects the views of the authors only and should not be construed to represent the views or policies of their affiliated organizations.} 
        
        \hspace{.4cm}
        Sandipan Samanta$^2$  
        
        \hspace{.4cm}
        Shoubhik Mondal$^3$  \vspace{.2cm}\\
        
   $^1$Daiichi-Sankyo Inc. (DSI), Basking Ridge, NJ, USA\\ 
   $^2$QIAGEN GmbH, Hilden, Germany\\
   $^3$Boehringer Ingelheim, CT, USA
}
\date{\today}

\begin{document}
\maketitle

\begin{abstract}
Assessment of study success using conditional power (CP), the predictive power of success (PPoS) and probability of success (PoS) is becoming increasingly common for resource optimization and adaption of trials in clinical investigation. Determination of these measures is often a non-trivial mathematical task. Further, the terminologies used across the literature are not consistent, and there is no consolidated presentation on this. Lastly, certain types of trials received more attention where others  (e.g., single-arm trial with time-to-event endpoints) were completely ignored.  We attempted to fill these gaps. This paper first provides a detailed derivation of CP, PPoS and PoS in a general setting with normally distributed test statistics and normal prior. Subsequently, expressions for these measures are obtained for continuous, binary, and time-to-event endpoints in single-arm and two-arm trial settings. We have discussed both clinical success and trial success. Importantly, we have derived the expressions for CP, PPoS and PoS in a single-arm trial with a time-to-event endpoint that was never addressed in the literature to our knowledge. In that discussion, we have also shown that commonly recommended $1/d$  consistently under-estimates the variance of log(median)  and alternative expression for variance was derived. We have also presented the PPoS calculation for the binomial endpoint with a beta prior. Examples are given along with the comparison of CP and PPoS. Expressions presented in this paper are implemented in \verb|LongCART| package in \verb|R|. An \verb|R| shiny app is also available at \url{https://ppos.herokuapp.com/}. 
\end{abstract}




\section{Introduction}\label{sec:intro}
The need to determine the probability of ``study success'' may arise at various stages of drug development. For example, the goal of such an exercise could be making a go or no go decision at the beginning of a prospective trial based on data from the earlier phase.  It can also be used to monitor an ongoing clinical trial to answer various questions: should the trial continue or should it stop? Is the current sample size sufficient? Do we need any adaption to the trial? The term ``success'' or ``study success'' is often understood in the context of achieving a pre-specified threshold for p-value (e.g., two-sided 0.05 or one-sided 0.025) at the end of the trial. In this paper, we would call it ``trial success'' to differentiate it from ``Clinical success''. ``Clinical success'' is defined as the observed treatment effect size exceeding some threshold value that is often clinically meaningful  \cite{dmitrienko2006bayesian, saville2014utility}. This paper focuses on the determination of the conditional power (CP), predictive power of success (PPoS) and probability of success (PoS) to assess the chance of ``trial success'' and ``clinical success''. \\

The {\it CP}, {\it PPoS} and {\it PoS} along with {\it power} are often used  statistical tools to quantify the chance of  success (either ``trial success'' or ``clinical success'').   Of these, power and PoS are calculated at the beginning of the trial, whereas CP and PPoS are determined after observing the interim results. All these measures attempt to determine pr(success) (i.e., chance of success) based on the prior belief of effect size at the beginning of the trial, or the observed value of effect size in the first part of the trial, or the combination of both. At the beginning of a trial, pr(success) is determined solely based on the prior belief. On the other hand, calculation of pr(success) at the interim analysis has to take following things into consideration: (a) available interim results, (b) limited uncertainty in the trial data arising from the post-interim part, and (c) the decision to use or not the prior information. Further,  power and CP are frequentist tools, whereas PoS and PPoS follow the Bayesian paradigm. The difference of Bayesian approach over the frequentist approach is the way available knowledge on effect size is summarized: Bayesian measures summarize this knowledge as distribution of effect whereas frequentist measure makes the best guess about the effect size as a single value, and thereby frequentist measures may not be a good indicator of pr(success) \cite{spiegelhalter1986predictive, o2005assurance, chuang2006sample, gillett1994average}. For this reason, PoS is also viewed as average `power' over the prior distribution of $\theta$ \cite{chuang2006sample, gillett1994average} whereas PPoS is the average CP over the predictive distribution of $\theta$. Further, in the Bayesian approach, both prior knowledge and available interim results can be used together in assessing pr(success); however, all available knowledge must be summarized into a single value in the frequentist approach.\\

CP is defined as pr(success) given the interim result and assuming a fixed effect size for the remainder of the trial. \citet{halperin1982aid} first proposed the use of CP in monitoring a long-term clinical trial. They proposed to calculate two CPs for trial success given the current results: (a) assuming the fixed effect size for the remainder of the trial as specified under the null hypothesis ($H_0$), and (b) assuming the fixed effect size as expected at the beginning of the trial under the alternative hypothesis ($H_1$). \citet{lan1988b} generalized the calculation of CP using the B-value (see Section~\ref{sec:prelim}). Further, \citet{lachin2005review}  has shown that any formal stopping boundaries based on the CP (along with the type I and II error probabilities) can be expressed using B value in a  study with interim futility analysis. The use of PPoS in a clinical trial can be traced back to \citet{choi1985early}: they employed the predictive distribution of proportion using a beta prior to obtaining the ``desired probability" of trial success. However, it was \citet{spiegelhalter1986monitoring} who first proposed a general Bayesian framework to obtain unconditional power by averaging the conditional probability of success over the {\it current opinion of the treatment effect} based on the results observed in the first part of the trial.  They termed this unconditional power as the ``predictive probability of rejecting the $H_0$.  In the remainder of this paper, this quantity is referred to as PPoS. For a prospective trial, \citet{spiegelhalter1986predictive} first suggested calculating ``average power''  as the ``overall predictive probability of obtaining a significant result",  where the so-called statistical power of the trial is averaged with respect over the prior distribution of belief about the possible effect size. To distinguish it from the PPoS based on interim data, we would call it PoS. The frameworks of \cite{choi1985early, spiegelhalter1986monitoring, spiegelhalter1986predictive} apply the Bayesian methodology in monitoring clinical trials where analyses were carried out using conventional frequentist techniques, and this framework was later referred to as `hybrid classical-Bayesian' \cite{spiegelhalter2004bayesian}.
\citet{lan2009conditional} have discussed the relationship between the CP and PPoS.  The majority of the earlier works (e.g., \cite{halperin1982aid, choi1985early, spiegelhalter1986monitoring}) considered a two-arm trial with a binary endpoint for illustration. However, these frameworks are illustrated for the continuous endpoints (e.g., see \cite{choi1989monitoring}) and the survival endpoints (e.g., see \cite{tang2015optimal}) as well. \\

Calculation of pr(success) (i.e., CP, PPoS and PoS) is often `non-trivial' mathematical task \cite{spiegelhalter1986predictive}. To make it more difficult, the literature in this area is severely suffered by the inconsistent use of terminologies for these concepts. For example, PPoS has been referred as  `predictive power' \cite{dmitrienko2006bayesian, lan2009conditional}, `Bayesian predictive power (BPP)' \cite{rufibach2016bayesian}, `predictive probability of statistical significance' \cite{saville2014utility} and `probability of study success' \cite{wang2013evaluating}  in the literature. On the other hand, the PoS in the literature is also referred in the literature as `average success probability \cite{chuang2006sample, gillett1994average}, `assurance' \cite{o2005assurance}, and `expected power'  \cite{gillett1994average}. Further, certain areas (e.g., trials with binary endpoint) received more attention compared to the rest. For example, we could not find any literature discussing these probability measures in a single-arm trial with the time-to-event endpoint. Despite the wide popularity of these measures, the current literature still lacks the concise presentation of these measures under a general framework of hypothesis testing. The present work attempts to fill all these gaps.\\

In this paper, we focus on CP, PPoS with or without prior and PoS. We first derive general expressions for these pr(success)  measures for normally distributed test statistics with normal prior  (Section~\ref{sec:genderive}) under a unified framework of hypothesis testing (Section~\ref{sec:prelim}).  Subsequently, we present the expressions for these measures in single-arm and two-arm trials with continuous, binary and time-to-event endpoints separately in Section~\ref{sec:byendpt}. For two-arm trials, examples are presented along with the comparison between CP and PPoS, and assessments of the impact of the prior distribution on predictive distribution of effect size and PPoS. Importantly, we have derived the expressions for CP, PPoS and PoS in a single-arm trial with the time-to-event endpoint in  Section~\ref{sec:survsingle} that was never addressed in the literature to our knowledge. In that discussion, we have also shown that commonly recommended approximated variance of $1/d$ (e.g. see \cite{brookmeyer1982confidence}) consistently under-estimates the variance of log(median) (see Figure~\ref{fig:median_sd}),  and we have derived alternative expression for variance. Expressions of PPoS for the binary endpoint with beta prior are presented in Section~\ref{sec:betabinom} with example. Implementation in R through LongCART package \cite{kundu2021longcart} and R shiny app (\url{https://ppos.herokuapp.com/}) are discussed in Section~\ref{sec:software} and illustrated in Appendix 2.

\section{Preliminaries and notations}\label{sec:prelim}

\begin{table}
\begin{center}
\caption{Summary of notations used in Section~\ref{sec:prelim} and Section~\ref{sec:genderive} in deriving general expression of CP, PPoS and PoS.} \label{tab:gennotation}
\begin{tabular} 
{|r|l|}
\hline
$t$  & fraction of information at interim $(0< t < 1)$. \\ \hline
$\theta$ & true treatment effect.\\\hline
$\hat{\theta}(t)$ & observed estimate of $\theta$ at the interim analysis with information $t$\\\hline
$\hat{\theta}(1-t)$ & projected value of $\theta$ from post-interim data with remaining information of $1-t$\\\hline
$k$ & SE of $\hat{\theta}(t)$ at final analysis ($t=1$)\\\hline
$\tilde{k}$ & projected SE in the trial\\\hline
$Z(t)$ & Z-test statistic based on interim data.\\\hline
$Z(1-t)$ & Z-test statistic based on post-interim data.\\\hline
$B(t)$ & B-values; $ B(t) = Z(t)\cdot \sqrt{t}$ \\ \hline
$c(t)$ & rejection boundary at analysis with information $t$ \\\hline
$\Phi(\cdot)$ & cumulative distribution function of a standard normal variate \\\hline
$\theta'$ & projected value of $\theta$ from post-interim data (excludes data upto interim).\\\hline
$\theta_0, \sigma_0$,  & mean and SD of a normal prior for $\theta$\\\hline
$\psi$ & $\sigma_0^2/(\sigma_0^2+k^2/t)$ \\\hline

$\theta_{\rm min}$ & threshold for ``clinical success" \\\hline
$\gamma$  &  $\gamma = c(1)$ for ``trial success" and $\gamma=\frac{\theta_{\rm min}}{k}$ for ``clinical success". \\\hline

\end{tabular}
\end{center}
\end{table}

Notations presented in this Section are listed in Table~\ref{tab:gennotation}. We consider the following general form of hypothesis testing in a clinical trial:
\[
H_0: \theta=0 \qquad {\rm vs.} \qquad H_1: \theta>0
\]
where $\theta$ is a parameter of interest.  For example, it could be either mean or proportion in a given population or their difference between two populations or log HR. We also assume results from the interim analysis performed after the accrual of $t$ amounts of ``information'' ($0\le t \le 1$) are available. At any given time of analysis, the information $t$ equals the proportion of interim subjects ($n$, say) to the maximum number of planned subjects ($N$, say) for continuous and binary endpoints, or proportion of observed events at interim analysis ($d$, say) to the maximum planned events ($D$, say) for time-to-event endpoints. Let's $\hat{\theta}(t)$  be the estimate of $\theta$ at the interim analysis with corresponding ndard error (SE) as $SE[\hat{\theta}(t)]=k\cdot \frac{1}{\sqrt{t}}$. Note that, $k$ is the SE of the estimate at the final analysis and does not depend on $t$. For example, $k=\sigma/\sqrt{N}$ in trials with continuous endpoint (where $\sigma$ is the standard deviation (SD)), and $k \approx 2/\sqrt{D}$ in trials with time-to-event endpoint, respectively. With this, for both Z-test and log-rank test, test statistic $Z(t)$ can be expressed as 
\[
Z(t)=\frac{\hat{\theta}(t)}{SE[\hat{\theta}(t)]} =\frac{\hat{\theta}(t)}{k} \cdot \sqrt{t}
\]
and,
\[
{\rm Reject}\;\; H_0 , \;\;\rm{ if } \;\; Z(t) > c(t)
\]
where $c(t)$ is the rejection boundary. $c(t)$ must be identified in advance and should be such that it preserves overall type I error of $\alpha$. In a single look design without any interim analysis, $c(1)=\Phi(1-\alpha)$, where $\Phi(\cdot)$ denotes the cumulative distribution function of a standard normal variate. In a multiple looks design with one or more interim analyses, $c(1)$ must be determined according to the appropriate alpha spending function (e.g. \cite{demets1994interim}).\\

Since, $E\left[Z(t)\right]= (\theta/k) \cdot \sqrt{t}$, the information growth in $Z(t)$ is proportional to $\sqrt{t}$. Following \citet{lan1988b}, the B-values are defined as follows:
\begin{equation}\label{eq:Bdef}
 B(t) = Z(t)\cdot \sqrt{t} = \frac{\hat{\theta}(t)}{k} \cdot t   
\end{equation}
with $B(0)=0$ at the trial initiation and $B(1)=Z(1)$ at the end of the trial. Further,
\begin{equation}\label{eq:var_B}
 {\rm Cov}[B(t), B(s)]=min(s,t) \qquad {\rm and} \qquad {\rm Var}[B(t)]=t   
\end{equation}
$B(t)$ has following advantages over $Z(t)$: (a) information growth in $B(t)$ is proportional to $t$ as $E[B(t)]=(\theta/k)\cdot t$, and (b) $B(t)$ has independent increments, implying $B(1) - B(t)$, is independent of the $B(t)$. Because of these two advantages, it is often easier to work with $B(t)$ compared to $Z(t)$.  \\

With interim data available at information time $t$, the uncertainty is now restricted to the results from the post-interim data (i.e., data contributing to remaining information of $(1-t)$). As $B(1)-B(t)$ is independent of $B(t)$, we can decompose $B(1)$ as follows
\[
B(1)= B(t) + [B(1) - B(t)]
\]
Based on Eq.~\eqref{eq:Bdef}, it translates to  
\begin{equation}\label{eq:z1}
Z(1)=\sqrt{t}\cdot Z(t) + \sqrt{1-t}\cdot Z(1-t)    
\end{equation}
where $Z(1-t)$ is the test statistic using the post-interim data only (i.e., data accrued after information $t$) and is defined as follows:
\begin{equation}\label{eq:z1minust}
 Z(1-t)=\frac{\hat{\theta}(1-t)}{SE\left[\hat{\theta}(1-t)\right]}
      =\frac{\hat{\theta}(1-t)}{k/\sqrt{1-t}}
      =\frac{\hat{\theta}(1-t)\cdot \sqrt{1-t}}{k}   
\end{equation}
where, $\hat{\theta}(1-t)$ is the estimate of $\theta$ based on post-interim  data only. From Eq.~\eqref{eq:z1}, we have
\begin{equation}\label{eq:theta1}
\hat{\theta}(1)=t\cdot \hat{\theta}(t) + (1-t)\cdot \hat{\theta}(1-t)
\end{equation}
Clearly,  $Z(1-t)$ and $\hat{\theta}(1-t)$ are independent from $Z(t)$ and $\hat{\theta}(t)$. Further, after the interim analysis, $Z(t)$ and $\hat{\theta}(t)$  are  fixed and known,  but $Z(1-t)$ and $\hat{\theta}(1-t)$ are  unknown and random. We call out ``Trial success'' at the time of final analyses if $Z(1)>c(1)$. Based on Eq.~\eqref{eq:z1} and Eq.~\eqref{eq:z1minust}, this translates to
\begin{equation*}
   \mbox{ Trial success, if  }\qquad    \hat{\theta}(1-t)  > \frac{k}{1-t} \cdot \left[c(1)-\sqrt{t}\cdot Z(t)\right]
\end{equation*}
Further, we call out ``clinical success'' at the time of final analyses if $\hat{\theta}(1)>\theta_{\rm min}$. Based on Eq.~\eqref{eq:theta1}, this translates to
\begin{equation*}
   \mbox{ Clinical success, if  }\qquad    \hat{\theta}(1-t)  > \frac{k}{1-t} \cdot \left[\frac{\theta_{\rm min}}{k}-\sqrt{t}\cdot Z(t)\right]
\end{equation*}
Note  the similarity in the definition of the ``Trial success'' and ``Clinical success'' criteria. By replacing $c(1)$ with $\frac{\theta_{\rm min}}{k}$ in the ``Trial success'' criterion, we can obtain the ``Clinical success'' criterion.  Therefore, we define the general criteria of ``success'' as follows:
\begin{equation}\label{eq:succ}
   \mbox{ Success, if  }\qquad    \hat{\theta}(1-t)  > \frac{k}{1-t} \cdot \left[\gamma-\sqrt{t}\cdot Z(t)\right]
\end{equation}
where $\gamma=c(1)$ for ``Trial success'' and $\gamma=\frac{\theta_{\rm min}}{k}$ for ``Clinical success''.

\section{Conditional power, Predictive power, Probability of success} \label{sec:genderive}


In this section, we first explain the concept of CP, PPoS and PoS.  Subsequently, the generic expressions for these measures are derived using normal approximation. Calculation of PPoS and PoS using Beta-Binomial distribution are presented in Section~\ref{sec:betabinom}.


\subsection{Conditional power (CP) based on interim results}

CP is the probability that the final study result will be statistically significant (or clinically successful), given the data observed thus far and a specific assumption about the pattern of the data to be observed in the remainder of the study, such as assuming the original design effect, or the effect estimated from the current data, or under the null hypothesis \cite{lachin2005review}.
Projecting the estimate of $\theta$ from the post-interim period to be $\theta'$,  the estimate of $\theta$ in the post-interim data would be distributed as
\[
    \hat{\theta}(1-t) \sim \mbox{Normal}\left[\theta' , \frac{k^2}{1-t}\right]
\]
Therefore, the CP is (from Eq~\eqref{eq:succ})
\begin{equation}\label{eq:cp}
    \mbox{CP}(t|\theta') = 1-\Phi\left(\frac{\frac{k}{1-t} \cdot \left[\gamma-\sqrt{t}\cdot Z(t)\right]-\theta'}{\frac{k}{\sqrt{1-t}}}\right)
    =\Phi\left(\frac{1}{\sqrt{1-t}}\left[\frac{\hat{\theta}(t)}{k}\left\{t+(1-t)\frac{\theta'}{\hat{\theta}(t)}\right\}-\gamma\right]\right) 
\end{equation}

It is very intuitive and common to replace $\theta'$ by $\hat{\theta}(t)$ in calculation of the CP. In that case, the expression of CP reduces to (e.g., see \cite{lan2009conditional})
\begin{equation} \label{eq:cp2}
    \mbox{CP}(t|\hat{\theta}(t))
    =\Phi\left(\frac{1}{\sqrt{1-t}}\left[ \frac{\hat{\theta}(t)}{k}-\gamma\right]\right) 
    =\Phi\left(\frac{1}{\sqrt{1-t}}\left[ \frac{Z(t)}{\sqrt{t}}-\gamma\right]\right) 
\end{equation}
This is the CP when the post-interim trend expected to follow the interim trend.


\subsection{Predictive power of success (PPoS) based on interim results}\label{sec:ppos}
CP depends on the specified treatment effect in the post-interim data, and therefore, calculation of CP can be arbitrary. An alternative would be to obtain PPoS as averaged CP over the predictive distribution of $\hat{\theta}(1-t)$ at interim. The use of prior in the calculation of PPoS optional. We have discussed PPoS below both with and without prior.\\  

Suppose the prior knowledge about $\theta$ can be summarized using the following prior distribution:
\begin{equation}\label{eq:prior}
\theta \sim {\rm Normal}\left[ \theta_0, \sigma_0^2\right]
\end{equation}
 With this prior, the posterior distribution of $\theta$ is
\[
\theta |\hat{\theta}(t)
    \sim 
    {\rm Normal}\left[ \psi \cdot \hat{\theta}(t) + (1-\psi)\theta_0, \;\;\; \psi \cdot k^2/t\right] 
\]
where, $\psi=\dfrac{\sigma_0^2}{\sigma_0^2+k^2/t}$ is the proportion of  the contribution of interim data. Accordingly, the predictive distribution of $\hat{\theta}(1-t)$ is as follows:
\[
\hat{\theta}(1-t)| \hat{\theta}(t) 
    \sim
    {\rm Normal}\left[\psi \cdot\hat{\theta}(t) + (1-\psi)\theta_0, \;\;\; k^2\left(\frac{1}{1-t}+\psi \cdot \frac{1}{t}\right)\right]
\]
We can now use the predictive distribution of $\hat{\theta}(1-t)$  and Eq.~\eqref{eq:succ} to derive the PPoS  as follows
\begin{align}
   \mbox{PPoS}(t|\mbox{interim}, \mbox{prior}) 
   &= 1-\Phi\left(
   \frac{\frac{k}{1-t} \left[\gamma-\sqrt{t}\cdot Z(t)\right]-\psi \cdot \hat{\theta}(t) - (1-\psi)\theta_0}
   {k \sqrt{1/(1-t)+\psi/t}}\right)\nonumber \\
       \label{eq:ppos}
\end{align}

This is the PPoS given the interim results and  the prior distribution. 
Without the prior distribution, PPoS can be derived as a special case of Eq~\eqref{eq:ppos} by setting $\psi=1$ which implies 100\% contribution of interim data to the predictive distribution of $\hat{\theta}(1-t)$. Therefore, after some simple algebraic manipulation, PPoS without the prior distribution is obtained as follows (e.g., see \cite{lan2009conditional, low2011perils}):
\begin{equation} \label{eq:PPoSwoPrior}
   \mbox{PPoS}(t|\mbox{interim}) 
    =\Phi\left(\frac{1}{\sqrt{1-t}}\left[\frac{Z(t)}{\sqrt{t}}-\gamma\right] \cdot \sqrt{t} \right) 
\end{equation}
Expressions of the PPoS in Eq.~\eqref{eq:PPoSwoPrior} and  the CP in Eq.~\eqref{eq:cp2} are very similar except the additional $\sqrt{t}$ inside $\Phi(\cdot)$ for PPoS. It's simple consequence is that CP$>$PPoS for  $CP>0.5$ and CP$<$PPoS for  $CP<0.5$ \cite{lan2009conditional} (see Figures~\ref{fig:example1}, \ref{fig:example2} and \ref{fig:example3}). That is, the CP is less extreme than the PPoS. For example, the stopping rule based on the PPoS will always make it harder to stop a trial compared to the CP.


\subsection{Probability of success (PoS) of  a prospective trial at the design stage}
The concept of {\it PoS} is very similar to {\it PPoS}: PPoS is averaged {\it CP} over the predictive distribution, whereas {\it PoS} is  averaged {\it power} over the predictive distribution. Unlike PPoS, PoS is calculated at the beginning of the trial, and hence it only relies on the prior distribution of $\theta$. As mentioned before, the PoS has also been referred to as `assurance' \cite{o2005assurance}, and `expected power' or `average success of probability' \cite{gillett1994average}. \\

With the prior distribution specified in Eq.~\eqref{eq:prior} and expecting the SE in the trial to be $\tilde{k}$, the predictive distribution of $\hat{\theta}(1)$ is
\[
\hat{\theta}(1)|\theta \sim {\rm Normal}\left[\theta_0, \;\;\; \sigma_0^2  + var[\hat{\theta}(1)]\right] \equiv {\rm Normal}\left[\theta_0, \;\;\; \sigma_0^2  + \tilde{k}^2\right]
\]
In that case, the PoS given the prior distribution would be
\begin{equation}\label{eq:pposbegin} 
   \mbox{PoS}=Pr[Z(1)>\gamma|\theta] 
    =Pr[\hat{\theta}(1)> \tilde{k}\cdot \gamma|\theta] 
    =\Phi\left(\frac{\theta_0 - \tilde{k}\cdot \gamma}{\sqrt{\sigma_0^2  + \tilde{k}^2}}\right) 
\end{equation}



\section{Expressions of CP, PPoS and PoS by type of endpoints }\label{sec:byendpt}

\begin{table}
\begin{center}
\caption{Summary of notations and expressions of CP, PPoS and PoS with normally distributed test statistics and normal priors presented in Section~\ref{sec:byendpt}} \label{tab:notation}
\resizebox{\textwidth}{!}{%
\begin{tabular}
{|l|c|c|c|c|c|c|}
\hline
& \multicolumn{2}{|c|} {Continuous endpoint}& \multicolumn{2}{|c|} {Binary endpoint} & \multicolumn{2}{|c|} {Time-to-event endpoint}\\\hline
& single-arm & two-arm & single-arm & two-arm & single-arm & two-arm\\\hline

 \multicolumn{7}{|l|} {}\\
 \multicolumn{7}{|l|} {Hypothesis and notations}\\\hline
Hypothesis & $H_0: \mu=\mu_1$ & $H_0: \mu_T-\mu_C=\Delta_1$
           & $H_0: \Pi=\Pi_1$ & $H_0: \Pi_T-\Pi_C=\Delta_1 $
           & $H_0: M=M_1$ & $H_0: \Delta=\Delta_1$\\
   & vs. & vs. & vs. & vs. & vs. & vs.\\
   & $H_1: \mu>\mu_1$ & $H_1: \mu_T-\mu_C>\Delta_1$
   & $H_1: \Pi>\Pi_1$ & $H_1: \Pi_T-\Pi_C>\Delta_1$ 
   & $H_1: M>M_1$ & $H_1: \Delta<\Delta_1$\\\hline
   
Timing of analyses &&&&&&\\ \hline
    \hspace{0.1in}$-$ interim analysis& $n$ subjects & $n$ subjects & $n$ subjects & $n$ subjects & $d$ events & $d$ events\\ \hline
    \hspace{0.1in}$-$ final analysis & $N$ subjects & $N$ subjects & $N$ subjects & $N$ subjects & $D$ events & $D$ events\\\hline

Allocation ratio & - & $a$ & - & $a$& - & $a$\\ \hline    

SD$^{1}$ & $\sigma$ & $\sigma$ & $\sigma=\sqrt{\Pi(1-\Pi)}$ & $\sigma$, $^2$see footnote  & - & -\\ \hline

Parameter of interest & mean ($\mu$) & mean diff ($\Delta$) & proportion ($\Pi$) & prop diff ($\Delta$) & median ($M$) & HR($\Delta$)\\
 &  & $\Delta=\mu_T - \mu_C$ &  & $\Delta=\Pi_T - \Pi_C$  &  & \\
  & & $\mu_T$: mean in Arm T& & $\Pi_T$: prop. in Arm T & &\\ 
 & & $\mu_C$: mean in Arm C& & $\Pi_C$: prop. in Arm C & &\\ \hline
  \hspace{0.1in}$-$value under $H_0$& $\mu_1$&$\Delta_1$ & $\Pi_1$ & $\Delta_1$& $M_1$&$\Delta_1$\\ \hline
 \hspace{0.1in}$-$interim estimate& $\bar{x}_n$&$\delta_n=\bar{x}_{T,n} - \bar{x}_{C,n}$ & $p_n$ & $\delta_n=p_{T,n} - p_{C,n}$& $m_d$&$\delta_d$\\ 
 & & $\bar{x}_{T,n}$: estimate of $\mu_T$& & $p_{T,n}$: estimate of $\Pi_T$ & &\\ 
 & & $\bar{x}_{C,n}$: estimate of $\mu_C$& & $p_{C,n}$: estimate of $\Pi_C$ & &\\ \hline
 \hspace{0.1in}$-$projected value for&&&&&&\\ 
 \hspace{0.2in}post-interim data& $\mu'$&$\Delta'$ & $\Pi'$ & $\Delta'$& $M'$&$\Delta'$\\ \hline
 \hspace{0.1in}$-$projected value in&&&&&&\\ 
 \hspace{0.2in}the entire trial& $\tilde{\mu}$&$\tilde{\Delta}$ & $\tilde{\Pi}$ & $\tilde{\Delta}$& $\tilde{M}$&$\tilde{\Delta}$\\ \hline
 \hspace{0.1in}$-$prior distribution& $\mu \sim$&$\Delta \sim$ & $\Pi \sim$ & $\Delta \sim$& $\log{M} \sim$&$\log{\Delta} \sim$\\ 
 \hspace{0.2in}& $N(\mu_0, \sigma_0^2)$&$N(\Delta_0, \sigma_0^2)$ & $N(\Pi_0, \sigma_0^2)$ & $N(\Delta_0, \sigma_0^2)$& $N(\log{M_0}, \sigma_0^2)$&$N(\log{\Delta_0}, \sigma_0^2)$\\\hline

Estimated (pooled) & $s_n$ & $s_n$ & $s_n$ & $s_n$ & - & - \\ 
SD at interim & &&&&&\\\hline
Projected (pooled) & $\tilde{\sigma}$ & $\tilde{\sigma}$ &  $\tilde{\sigma}=\sqrt{\tilde{\Pi}(1-\tilde{\Pi})}$ & $\tilde{\sigma}$  & - & - \\ 
SD at beginning & &&&&&\\\hline

 \multicolumn{7}{|l|} {}\\
 \multicolumn{7}{|l|} {At the beginning of trial}\\\hline
 \hspace{0.1in}$-$probability of&&&&&&\\ 
 \hspace{0.2in}success (PoS) 
 & Eq.~\eqref{eq:pp3contsingle}&Eq.~\eqref{eq:pp3cont} 
 & Eq.~\eqref{eq:pp3binarysingle}&Eq.~\eqref{eq:pp3cont}$^3$ & Eq.~\eqref{eq:pp3survsingle}&Eq.~\eqref{eq:pp3surv} \\ \hline
 
 \multicolumn{7}{|l|} {}\\
 \multicolumn{7}{|l|} {At the interim analysis}\\\hline
 \hspace{0.1in}$-$Conditional power&&&&&&\\ 
 \hspace{0.2in}(CP) with specified
 & Eq.~\eqref{eq:cpcontsingle}&Eq.~\eqref{eq:cpcont} 
 & Eq.~\eqref{eq:cpbinarysingle}&Eq.~\eqref{eq:cpcont}$^3$ & Eq.~\eqref{eq:cpsurvsingle}&Eq.~\eqref{eq:cpsurv} \\ 
 \hspace{0.2in}trend&&&&&&\\\hline
 \hspace{0.1in}$-$Conditional power&&&&&&\\ 
 \hspace{0.2in}(CP) with interim
 & Eq.~\eqref{eq:cp2contsingle}&Eq.~\eqref{eq:cp2cont} 
 & Eq.~\eqref{eq:cp2binarysingle}&Eq.~\eqref{eq:cp2cont}$^3$ & Eq.~\eqref{eq:cp2survsingle}&Eq.~\eqref{eq:cp2surv} \\ 
 \hspace{0.2in}trend&&&&&&\\\hline
 \hspace{0.1in}$-$predictive power&&&&&&\\ 
 \hspace{0.2in}of success (PPoS) 
 & Eq.~\eqref{eq:ppcontsingle}&Eq.~\eqref{eq:ppcont} 
 & Eq.~\eqref{eq:ppbinarysingle}&Eq.~\eqref{eq:ppcont}$^3$ & Eq.~\eqref{eq:ppsurvsingle}&Eq.~\eqref{eq:ppsurv} \\ 
 \hspace{0.2in}without prior&&&&&&\\ \hline 
 \hspace{0.1in}$-$predictive power&&&&&&\\ 
 \hspace{0.2in}of success (PPoS) 
 & Eq.~\eqref{eq:pp2contsingle}&Eq.~\eqref{eq:pp2cont} 
 & Eq.~\eqref{eq:pp2binarysingle}&Eq.~\eqref{eq:pp2cont}$^3$ & Eq.~\eqref{eq:pp2survsingle}&Eq.~\eqref{eq:pp2surv} \\ 
 \hspace{0.2in}with prior&&&&&&\\ \hline 
 \multicolumn{7}{l} {$-$ Subscripts $T$ and $C$ indicates measurements in treatment (T) and control (C) arms, respectively.}\\
 \multicolumn{7}{l} {$-$ $^1$SD: Standard deviation. For two arm trials it is the pooled SD.}\\
 \multicolumn{7}{l} {$-$ $^2$SD in two-arm trial with binary endpoint: $\sigma=\sqrt{\dfrac{a}{a+1}\left\{\dfrac{\Pi_T(1-\Pi_T)}{a}+\Pi_C(1-\Pi_C)\right\}}$}\\
 \multicolumn{7}{l} {$-$ $^3$The expression for PoS, CP and PPoS are identical for two-arm trials with binary and continuous endpoints, although the }\\
 \multicolumn{7}{l} {quantities (e.g., $\delta_n$, $s_n$ etc.) included in the expression  are to be obtained differently.}\\
\end{tabular}}
\end{center}
\end{table}

Expressions of CP, PPoS and PoS for  continuous, binary and survival endpoints in a single-arm and two-arm trials are derived separately in this section  based on the general expressions presented in the previous section with normal approximation. Summary of notations and expressions of CP, PPoS and PoS discussed in the Section are presented are presented in Table~\ref{tab:notation}. 
For the two-arm trial, the allocation ratio (treatment arm to control arm) is denoted as $a:1$. We denote $r^2=(a+1)^2/a$. The expressions for single-arm trial can be obtained directly from the corresponding expressions from the two-arm trial by specifying $r=1$. Intuition for setting $r=1$ is simple: the single-arm design can be thought of as $1:0$ allocation ratio (instead of $a:1$) in  which case, $r=(1+0)/\sqrt{1} = 1$. For clarity, we present the expressions for both single-arm and two-arm scenario separately. \\

We also would like to remind that here the expressions are presented for general success criteria  as presented in Eq.~\eqref{eq:succ}. One needs to set $\gamma=c(1)$ for ``Trial success'' and $\gamma=\frac{\theta_{\rm min}}{k}$ for ``Clinical success''. We have illustrated the calculation of CP, PPoS and PoS based on published clinical trial results for two-arm trials and also compared the behaviour of CP and PPoS. 


\subsection{Continuous endpoint, single-arm trial}

We start with the single-arm trial continuous endpoint.  Denote the population mean as $\mu$ and the maximum sample size in the trial as $N$.  We test the following hypotheses:
\[
H_0: \mu=\mu_1 \qquad {\rm vs.} \qquad H_1: \mu>\mu_1
\]
Here, $\theta =\mu-\mu_1$. At interim analysis with sample size $n$, the estimate of $\theta$ is $\hat{\theta}(t)=\overline{x}_n- \mu_1$ where $\overline{x}_n$ is the sample mean at interim. The corresponding test statistic is
\[
Z(t)=\frac{\overline{x}_n-\mu_1}{s_n/\sqrt{n}}=\frac{(\overline{x}_n-\mu_1)\sqrt{n}}{s_n}
\]
where $s_n$ is the estimate of SD ($\sigma$) at interim analysis. Further, in this case, $t=n/N$ and $k=s_n /\sqrt{N}$.\\ 

\textsl{Conditional power (CP)}: The CP with the future trend of the sample mean as $\mu'$ is
\begin{equation} \label{eq:cpcontsingle} 
       \Phi\left(\frac{1}{s_n}\sqrt{\frac{N}{N-n}}\left[\frac{1}{\sqrt{N}}\{n(\bar{x}_n-\mu_1)+(N-n)(\mu'-\mu_1)\}-s_n \cdot \gamma\right]\right)
\end{equation}
If we assume that the current trend observed through interim analysis continues to hold for future data as well (i.e., $\mu'=\overline{x}_n$), then the expression of the CP reduces to
\begin{equation} \label{eq:cp2contsingle} 
   \Phi\left(\frac{1}{s_n}\sqrt{\frac{N}{N-n}}\left[(\bar{x}_n-\mu_1)\sqrt{N}-s_n \cdot \gamma\right]\right)
\end{equation}

\textsl{Predictive power of success (PPoS)}: The PPoS solely based on the interim information can be expressed as
\begin{equation}
       \Phi\left(\frac{1}{s_n}\sqrt{\frac{n}{N-n}}\left[(\bar{x}_n-\mu_1)\sqrt{N}-s_n \cdot \gamma\right]\right)  \label{eq:ppcontsingle} 
\end{equation}
Incorporating prior information specified in Eq.~\eqref{eq:priorcont2}, expression of PPoS can be refined as
\begin{equation} \label{eq:pp2contsingle} 
    \Phi\left(
    \frac{1}{s_n}\sqrt{\frac{n}{N-n}}
    \frac
    {\left[(1-\psi)\{n(\bar{x}_n - \mu_1)+(N-n)(\mu_0 - \mu_1)\}/\sqrt{N} + \psi (\bar{x}_n - \mu_1) \sqrt{N} - s_n \gamma \right]}
    {\sqrt{\psi+(1-\psi)n/N}}\right)
\end{equation}
where, $\psi=n \sigma_0^2/(n \sigma_0^2+ s_n^2)$.\\

\textsl{Probability of success (PoS)}: The PoS of a prospective clinical trial with $N$ subjects and prior information specified in Eq.~\eqref{eq:priorcont2} can be expressed as 
\begin{equation}
        \Phi \left( \frac{\sqrt{N}\cdot (\mu_0 - \mu_1) - \tilde{\sigma} \cdot \gamma}{\sqrt{N \cdot \sigma_0^2 + \tilde{\sigma}^2}} \right) 
    \label{eq:pp3contsingle} 
\end{equation}
where $\tilde{\sigma}$ is the projected SD and $\tilde{k}=\tilde{\sigma}/\sqrt{N}$ is the projected SE in the trial.  For the calculation of PPoS with prior distribution in Eq.~\eqref{eq:pp2contsingle} and PoS in Eq.~\eqref{eq:pp3contsingle}, following prior for $\mu$ was used
\begin{equation}\label{eq:priorcont2}
    \mu  \sim  {\rm Normal}\left[ \mu_0, \sigma_0^2 \right]
\end{equation}

\subsection{Continuous endpoint, two-arm trial}\label{sec:conttwo}
Now consider a two-arm trial comparing treatment (T) with control (C) arm with population means as $\mu_T$ and $\mu_C$, respectively. Denote the maximum total sample size as $N$. We  test
\[
H_0: \mu_T-\mu_C=\Delta_1 \qquad {\rm vs.} \qquad H_1: \mu_T-\mu_C>\Delta_1
\]
We have, $\theta =\mu_T - \mu_C - \Delta_1$. At interim analysis with total sample size $n$,  $\hat{\theta}(t)=\delta_n - \Delta_1$ where $\delta_n=\bar{x}_{T,n}-\bar{x}_{C,n}$,  and $\bar{x}_{T,n}$ and $\bar{x}_{C,n}$ are the sample means at interim in treatment and control arms, respectively. The corresponding test statistic is
\[
Z(t)=\frac{\delta_n-\Delta_1}{r\cdot s_n/\sqrt{n}}=\frac{(\delta_n-\Delta_1)\sqrt{n}}{r\cdot s_n}
\]
where $s_n$ is the estimate of pooled SD ($\sigma$) at interim analysis. Further, $t=n/N$ and $k=r\cdot s_n /\sqrt{N}$.\\ 

\textsl{Conditional power (CP)}: The CP with estimated mean difference from post-interim data as $\Delta'$ is
\begin{equation}\label{eq:cpcont}
       \Phi\left(\frac{1}{r\cdot s_n}\sqrt{\frac{N}{N-n}}\left[\frac{1}{\sqrt{N}}\{n(\delta_n-\Delta_1)+(N-n)(\Delta' - \Delta_1)\}-r \cdot s_n \cdot \gamma\right]\right)
\end{equation}
Assuming the interim trend continues to hold for the future data (i.e., $\Delta'=\delta_n$), the CP reduces to
\begin{equation}\label{eq:cp2cont}
   \Phi\left(\frac{1}{r\cdot s_n}\sqrt{\frac{N}{N-n}}\left[(\delta_n-\Delta_1)\sqrt{N}-r \cdot s_n \cdot \gamma\right]\right)
\end{equation}

\textsl{Predictive Power of success (PPoS)}: The PPoS without  prior distribution can be expressed as
\begin{equation}
        \Phi\left(\frac{1}{r\cdot s_n}\sqrt{\frac{n}{N-n}}\left[(\delta_n-\Delta_1)\sqrt{N}-r\cdot s_n \cdot \gamma\right]\right) \label{eq:ppcont}
\end{equation}
The PPoS incorporating prior information specified in Eq.~\eqref{eq:priorcont} can be expressed as
\begin{equation} \label{eq:pp2cont} 
    \Phi\left(
    \frac{1}{rs_n}\sqrt{\frac{n}{N-n}}
    \frac
    {\left[(1-\psi)\{n(\delta_n - \Delta_1)+(N-n)(\Delta_0 - \Delta_1)\}/\sqrt{N} + \psi (\delta_n - \Delta_1) \sqrt{N} - r \cdot s_n \gamma \right]}
    {\sqrt{\psi+(1-\psi)n/N}}\right)
\end{equation}
where, $\psi=n \sigma_0^2/(n \sigma_0^2+r^2 s_n^2)$.\\

\textsl{Probability of success (PoS)}: The PoS of a prospective clinical trial with $N$ subjects and prior information specified in Eq.~\eqref{eq:priorcont} can be expressed as (e.g. see \cite{o2005assurance})
\begin{equation}
        \Phi \left( \frac{\sqrt{N}\cdot (\Delta_0 - \Delta_1) -  r \cdot \tilde{\sigma} \cdot \gamma}{\sqrt{N\cdot\sigma_0^2 + r^2 \cdot \tilde{\sigma}^2}} \right)   \label{eq:pp3cont} 
\end{equation}
where $\tilde{\sigma}$ is the projected pooled SD and $\tilde{k}=r\cdot\tilde{\sigma}/\sqrt{N}$ is the projected SE in the trial. For PPoS with prior distribution (Eq.~\eqref{eq:pp2cont}) and PoS (Eq.~\eqref{eq:pp3cont}), following prior was used
\begin{equation}\label{eq:priorcont}
    \mu_T - \mu_C  \sim  {\rm Normal}\left[ \Delta_0, \sigma_0^2 \right]
\end{equation}

\textbf{Example 1:} In the pragmatic, unblinded, non-inferiority CODA trial \cite{coda2020randomized}, 1552 subjects (=$N$) with appendicitis were equally randomized to receive either antibiotics or to undergo appendectomy.  The primary outcome was 30-day health status, as assessed with the European Quality of Life–5 Dimensions (EQ-5D) questionnaire (scores range from 0 to 1, with higher scores indicating better health status; non-inferiority margin, 0.05 points). For this illustration, we imagine to have an interim analysis at the sample size of 776 (=$n$). According to O'Brien alpha spending function, the  rejection boundaries for Z test statistic are 2.96 and 1.97 (=$c(1)$) at interim and final analyses, respectively.\\

\begin{figure}
\centering
\caption{{\it Left:} Predictive distributions of $\Delta=\mu_T - \mu_C$ and {\it Right:} plots of the CP and PPoS for trial success against $\delta_n$ (interim estimate of $\Delta$) for Example 1. Horizontal and vertical reference lines in the left panel correspond to 50\% power and observed value of $\delta_n=-0.025$, respectively.}
\label{fig:example1}
\begin{subfigure}{.49\textwidth}
  \centering
  \includegraphics [angle=0,width=80mm, height=80mm]{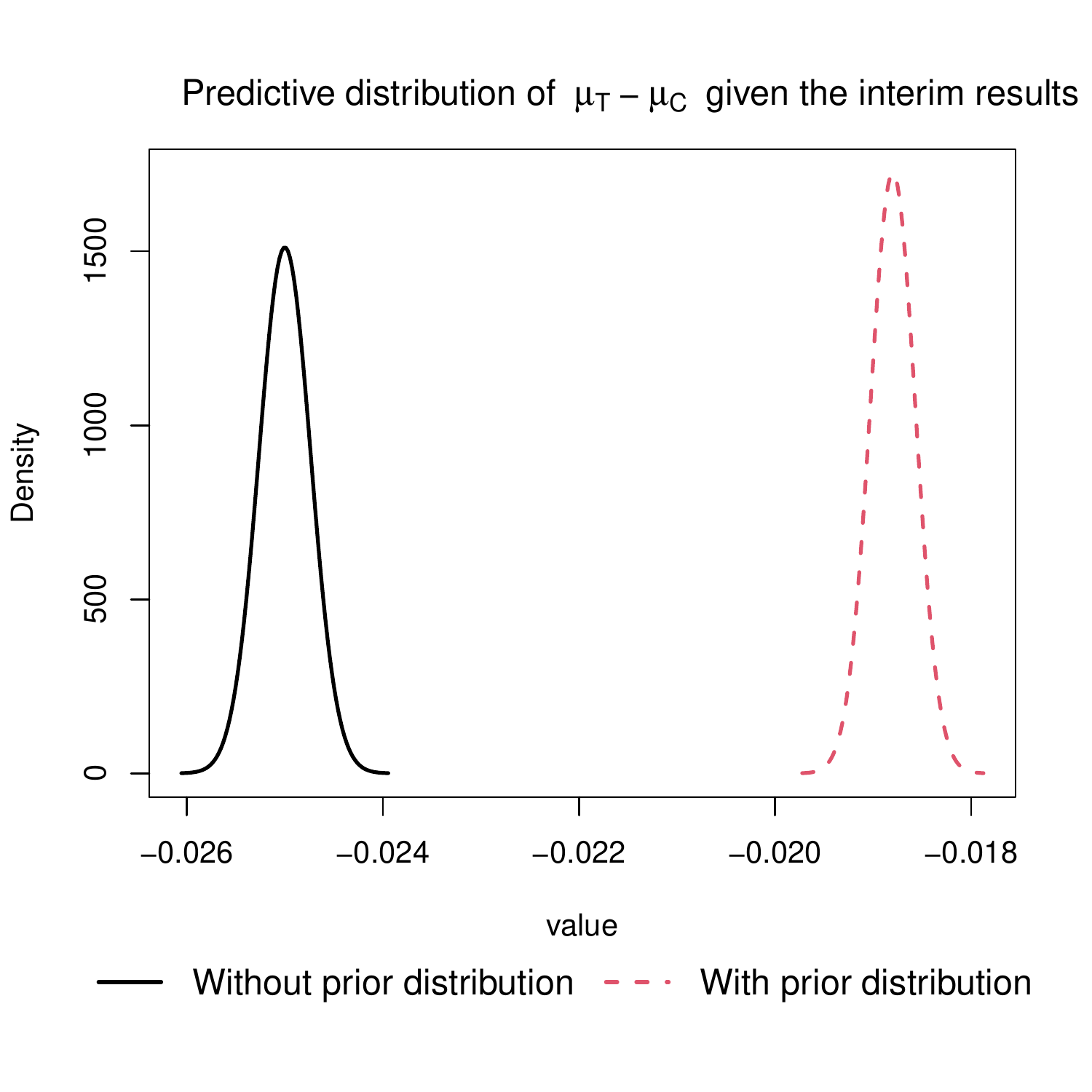}
\end{subfigure}
\begin{subfigure}{.49\textwidth}
  \centering
\includegraphics [angle=0,width=80mm, height=80mm]{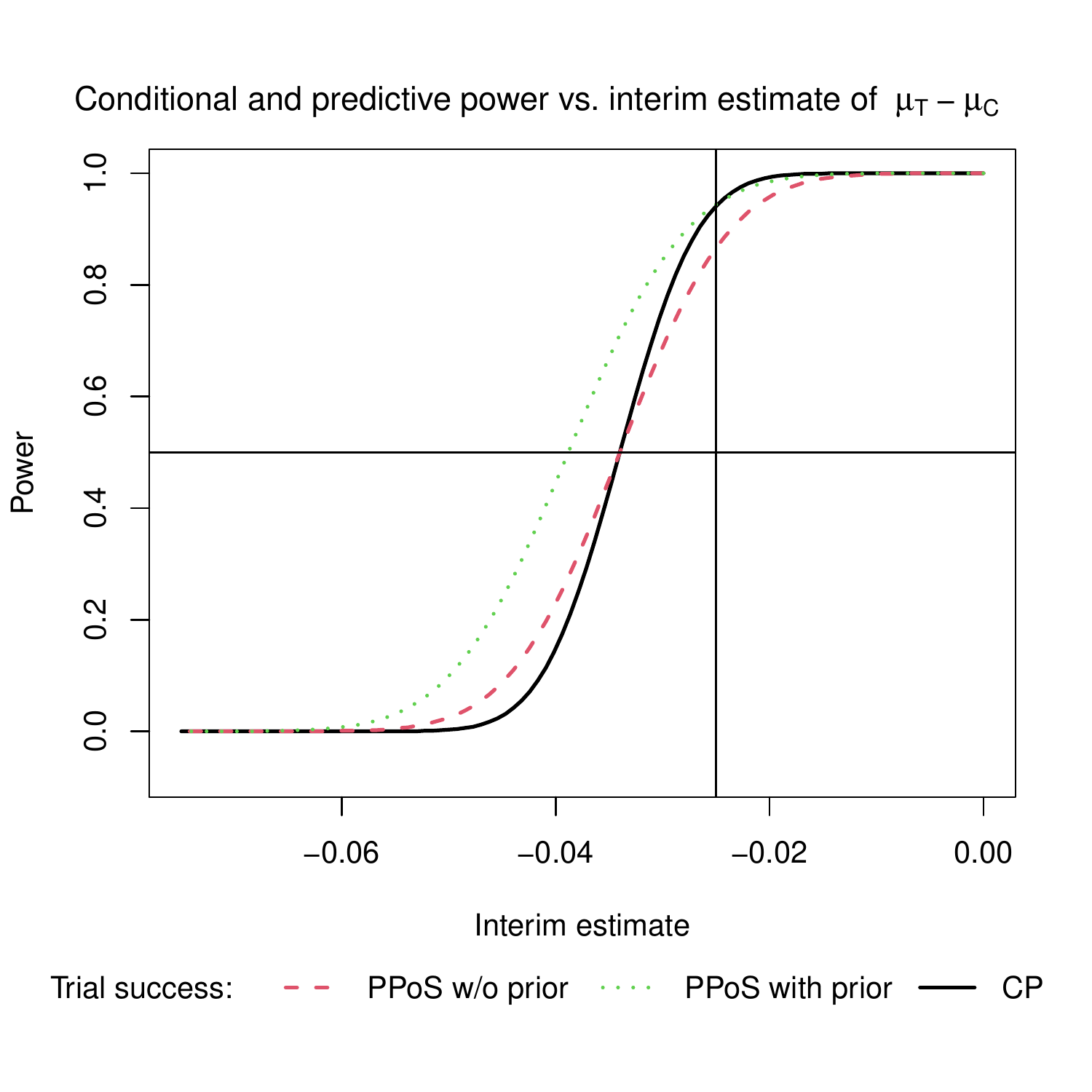}
\end{subfigure}
\end{figure}

We are testing following hypotheses: $H_0: \mu_T - \mu_C \le -0.05$ vs. $H_1: \mu_T - \mu_C>-0.05$. Therefore, $\Delta_1=-0.05$. Further, the external information  are summarized as following prior distribution 
\[
    \mu_T - \mu_C  \sim  {\rm Normal}\left[ \Delta_0=0, \;\;\sigma_0^2=(0.02)^2 \right]
\]
Treatment allocation ratio is 1:1; therefore, $r^2=(1+1)^2/2=4$. First, we illustrate the calculation of the PoS at the design stage: If the projected pooled SD in the trial was 0.12 ($=\tilde{\sigma}$), the PoS for trial success ($\gamma=c(1)=1.97$)  would be $\Phi\left(\frac{\sqrt{1552}(0-(-0.05))-(2)(0.12)(1.97)}{\sqrt{(1552)(0.02)^2+(4)(0.12)^2}}\right)=0.965$  (see Eq.~\eqref{eq:pp3cont}). \\

For the calculation of CP and PPoS, let's consider following interim results: mean difference of -0.025 (=$\delta_n$) points with SD as 0.16 (=$s_n$). 
Assuming the interim trend to be continued to the remaining part of trial as well, based on Eq.~\eqref{eq:cp2cont}, the conditional power for trial success ($\gamma=c(1)=1.97$) is 0.941 and PPoS would be 0.866 (see Eq.~\eqref{eq:pp2cont}).  Now, expecting -0.030 mean difference from post-interim data (i.e., $\Delta'=-0.030$), the conditional power would be 0.871. Further,  the PPoS for trial success  given the interim results and prior distribution is 0.944 (see Eq.~\eqref{eq:ppcont}). \\

The predictive distributions of $\Delta=\mu_T -\mu_C$  with and without prior distribution,  and the CP and PPoS values for trial success against $\delta_n$ (i.e., interim estimate of mean difference) are plotted in Figure~\ref{fig:example1}.  As mentioned in Section~\ref{sec:ppos}, we can verify that CP$>$PPoS for  $CP>0.5$ and CP$<$PPoS for  $CP<0.5$, when prior information was not incorporated in PPoS. Both the predictive distribution and PPoS were improved due to use of optimistic prior.

\subsection{Binary endpoint, single-arm trial} \label{sec:binarysingle}
Let's $\Pi$ denotes the population proportion in a single-arm trial  with the binary endpoint. The maximum sample size in the study is $N$. We test the following set of hypotheses:
\[
H_0: \Pi=\Pi_1 \qquad {\rm vs.} \qquad H_1: \Pi>\Pi_1
\]
We have, $\theta =\Pi- \Pi_1$. At interim analysis with sample size $n$, $\hat{\theta}(t)=p_n - \Delta_1$ with $p_n$ being the sample proportion. The corresponding test statistic is
\[
Z(t)=\frac{p_n-\Pi_1}{s_n/\sqrt{n}}=\frac{(p_n-\Pi_1)\sqrt{n}}{s_n}
\]
where, $s_n=\sqrt{p_n(1-p_n)}$ is the estimate of SD ($\sigma$). Further, $t=n/N$ and $k=s_n /\sqrt{N}={\rm SE}(p_n) \cdot \sqrt{t}$. Expressions of the CP, PPoS and PoS in this case can be obtained from the corresponding expressions in single-arm trial with continuous endpoint by replacing $\bar{x}_n$ with $p_n$, $\mu_1$ with $\Pi_1$, and $\mu'$ with $\Pi'$.\\

\textsl{Conditional power (CP)}: The CP with projected proportion of $\Pi'$ for post-interim data is
\begin{equation} \label{eq:cpbinarysingle} 
       \Phi\left(\frac{1}{s_n}\sqrt{\frac{N}{N-n}}\left[\frac{1}{\sqrt{N}}\{n(p_n-\Pi_1)+(N-n)(\Pi'-\Pi_1)\}-s_n \cdot \gamma\right]\right)
\end{equation}
If the interim trend continues to hold for the future data (i.e., $\Pi'=p_n$), the CP reduces to
\begin{equation} \label{eq:cp2binarysingle} 
   \Phi\left(\frac{1}{s_n}\sqrt{\frac{N}{N-n}}\left[(p_n-\Pi_1)\sqrt{N}-s_n \cdot \gamma\right]\right)
\end{equation}

\textsl{Predictive power of success (PPoS)}: The PPoS based on the interim information can be expressed as
\begin{equation}
       \Phi\left(\frac{1}{s_n}\sqrt{\frac{n}{N-n}}\left[(p_n-\Pi_1)\sqrt{N}-s_n \cdot \gamma\right]\right)  \label{eq:ppbinarysingle} 
\end{equation}
Incorporating prior information specified in Eq.~\eqref{eq:priorbinarysingle}, expression of the PPoS can be refined as
\begin{equation} \label{eq:pp2binarysingle} 
    \Phi\left(
    \frac{1}{s_n}\sqrt{\frac{n}{N-n}}
    \frac
    {\left[(1-\psi)\{n(p_n - \Pi_1)+(N-n)(\Pi_0 - \Pi_1)\}/\sqrt{N} + \psi (p_n - \Pi_1) \sqrt{N} - s_n \gamma \right]}
    {\sqrt{\psi+(1-\psi)n/N}}\right)
\end{equation}
where, $\psi=n \sigma_0^2/(n \sigma_0^2+N s_n^2)$.\\

\textsl{Probability of success (PoS)}: The PoS of a prospective clinical trial with $N$ subjects and prior information specified in Eq.~\eqref{eq:priorbinarysingle} can be expressed as 
\begin{equation}
        \Phi \left( \frac{\sqrt{N}\cdot (\Pi_0 - \Pi_1) - \tilde{\sigma} \cdot \gamma}{\sqrt{N \cdot \sigma_0^2 + \tilde{\sigma}^2}} \right) 
    \label{eq:pp3binarysingle} 
\end{equation}
where, $\tilde{\sigma}=\sqrt{\tilde{\Pi}(1-\tilde{\Pi})}$ is the projected SD and $\tilde{k}=\tilde{\sigma}/\sqrt{N}$ is the projected SE in the trial with $\tilde{\Pi}$ being the projected proportion in the trial. Following prior was used for the PPoS with prior distribution in Eq.~\eqref{eq:pp2binarysingle} and PoS in Eq.~\eqref{eq:pp3binarysingle}
\begin{equation}\label{eq:priorbinarysingle}
    \Pi  \sim  {\rm Normal}\left[ \Pi_0, \sigma_0^2=\sigma_0^2 \right]
\end{equation}


\subsection{Binary endpoint, two-arm trial}\label{sec:binarytwo}
Consider a two-arm trial comparing treatment (T) with control (C) arm with population proportions as $\Pi_T$ and $\Pi_C$, respectively. Denote the maximum total sample size as $N$. We  test
\[
H_0: \Pi_T-\Pi_C=\Delta_1 \qquad {\rm vs.} \qquad H_1: \Pi_T-\Pi_C>\Delta_1
\]
Here, $\theta =\Pi_T-\Pi_C - \Delta_1$. At interim analysis with total sample size $n$,  $\hat{\theta}(t)=\delta_n - \Delta_1$ where $\delta_n=p_{T,n}-p_{C,n}$ is the difference between the estimated proportions ($p_{T,n}$ and $p_{C,n}$) in the two arms. 
\[
SE(\delta_n)=\sqrt{\dfrac{p_{T,n}(1-p_{T,n})}{a\cdot n/(1+a)}+\dfrac{p_{C,n}(1-p_{C,n})}{n/(1+a)}}=r \cdot s_n/\sqrt{n}
\]
where $s_n^2=\dfrac{a}{a+1}\left\{\dfrac{p_{T,n}(1-p_{T,n})}{a}+p_{C,n}(1-p_{C,n})\right\}$ is the estimate of pooled SD ($\sigma$). Therefore, the corresponding test statistic is
\[
Z(t)=\frac{\delta_n-\Delta_1}{r \cdot s_n/\sqrt{n}}=\frac{(\delta_n-\Delta_1)\sqrt{n}}{r\cdot s_n}
\]
Further, $t=n/N$ and $k=r\cdot s_n /\sqrt{N}={\rm SE}(\delta_n) \cdot \sqrt{t}$. Expressions of the CP, PPoS and PoS for two-arm trial with binary endpoint are same with that of continuous case in Section~\ref{sec:conttwo}. That is, 
\begin{itemize}
    \item Eq.~\eqref{eq:cpcont} for the CP with the projected difference from post-interim data as $\Delta'$.
    \item Eq.~\eqref{eq:cp2cont} for the CP with the projected difference similar to that observed at interim analysis.
    \item Eq.~\eqref{eq:ppcont} for the PPoS without prior distribution.
    \item Eq.~\eqref{eq:pp2cont} for the PPoS with prior distribution with $\psi=n \sigma_0^2/(n \sigma_0^2+N s_n^2)$.
    \item Eq.~\eqref{eq:pp3cont} for the PoS with projected SD as $\tilde{\sigma}=\sqrt{\dfrac{a}{a+1}\left\{\dfrac{\tilde{\Pi}_T(1-\tilde{\Pi}_T)}{a}+\tilde{\Pi}_C(1-\tilde{\Pi}_C)\right\}}$ and the projected SE as $\tilde{k}=r\cdot \tilde{\sigma}/\sqrt{N}$,  where $\tilde{\Pi}_T$ and $\tilde{\Pi}_C$ are the projected proportions in the trial.
\end{itemize}
For the PPoS with prior distribution and PoS, following prior for $\Pi_T - \Pi_C$ was considered
\begin{equation}\label{eq:priorbinary}
    \Pi_T - \Pi_C  \sim  {\rm Normal}\left[ \Delta_0, \sigma_0^2=\sigma_0^2 \right]
\end{equation}

\textbf{Example 2:} \citet{fenaux2020luspatercept} reported the trial results of placebo-controlled, phase 3 trial evaluating the effect of Luspatercept in patients with lower-risk myelodysplastic syndromes. The primary endpoint was the proportion of patients with transfusion independence for eight weeks or longer during weeks 1 through 24. A total sample size of 210 patients (=$N$) with 2:1 treatment allocation ratio would give the study 90\% power to detect differences between response rates of 0.30 in the luspatercept arm and 0.10 in the placebo arm with the one-sided alpha of 0.025 and 10\% dropout rate. For this illustration, we add an interim analysis at the sample size of 158 (=$n$). Further, we assume the clinically meaningful difference is  15\% (=$\theta_{\min}$). According to O'Brien alpha spending function, the  rejection boundaries for the Z test statistic are 2.34  and 2.012 (=$c(1)$) at interim and final analyses, respectively. \\

\begin{figure}
\centering
\caption{{\it (Left)} Predictive distributions of $\Delta=\Pi_T - \Pi_C$, and {\it (Right)} plots of the CP and PPoS for trial success against $\delta_n$ (interim estimate of $\Delta$) for Example 2. Horizontal and vertical reference lines in the left panel correspond to 50\% power and observed value of $\delta_n=0.157$, respectively.}
\label{fig:example2}
\begin{subfigure}{.49\textwidth}
  \centering
  \includegraphics [angle=0,width=80mm, height=80mm]{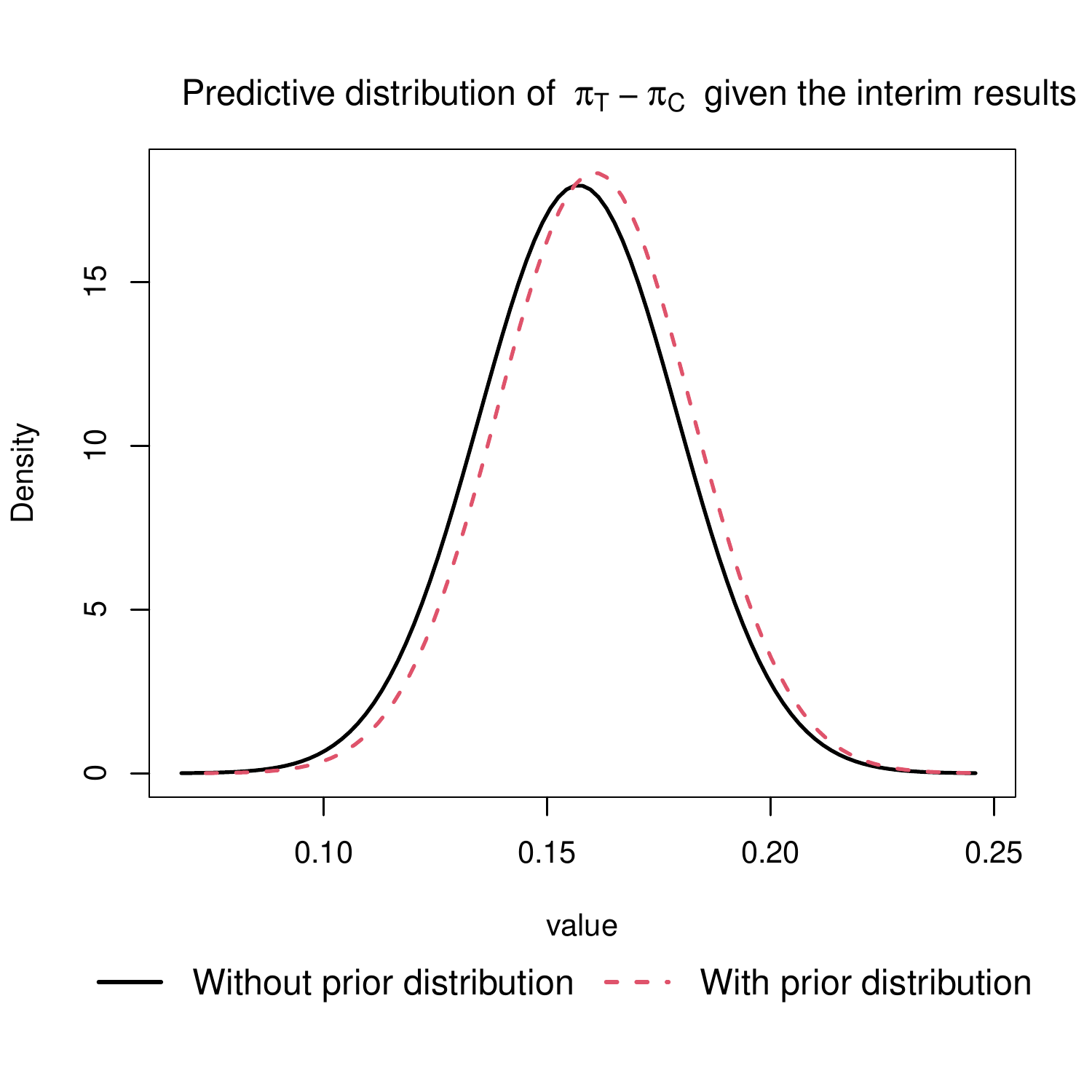}
\end{subfigure}
\begin{subfigure}{.49\textwidth}
  \centering
\includegraphics [angle=0,width=80mm, height=90mm]{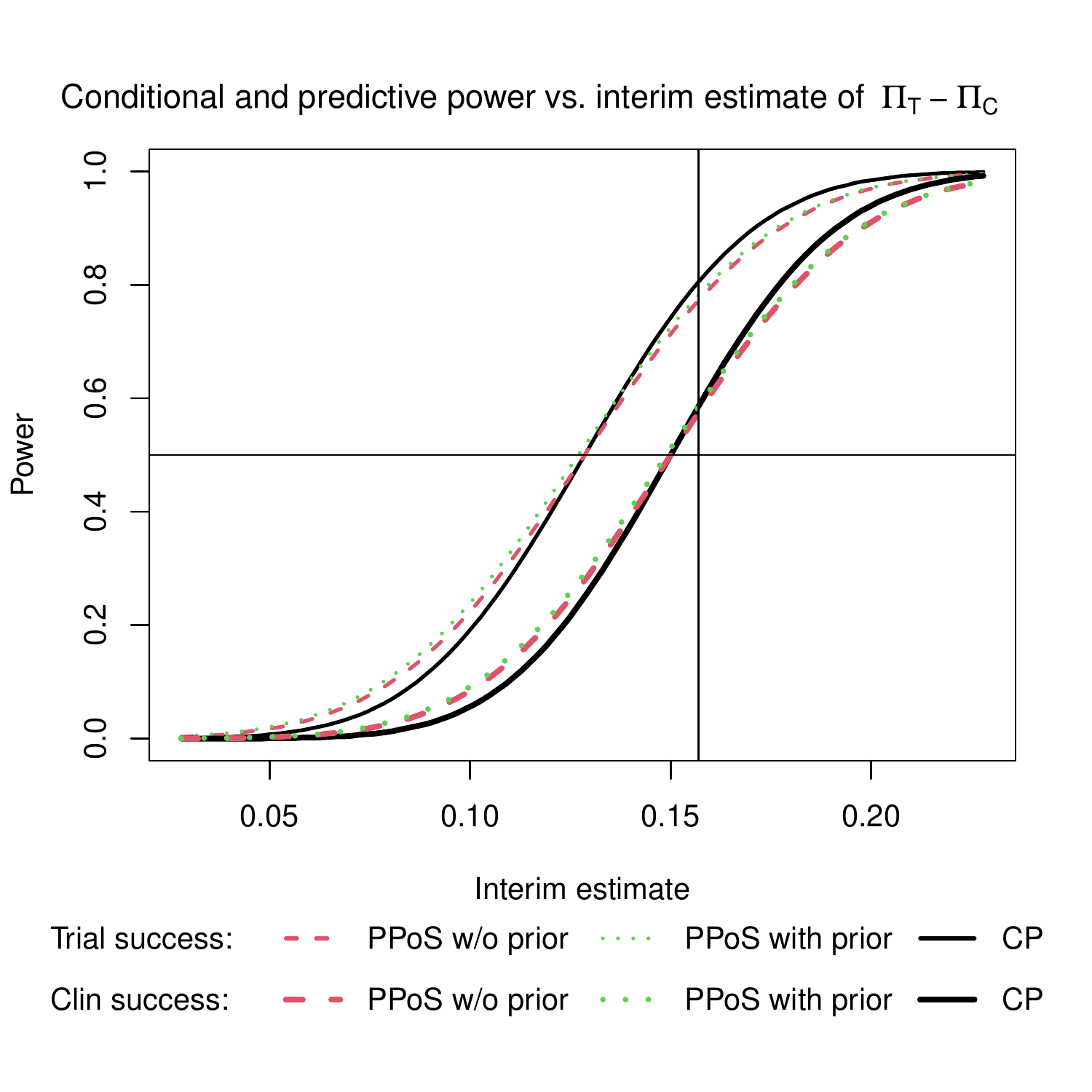}
\end{subfigure}
\end{figure}

In this case, we are statistically testing the following hypotheses: $H_0: \Pi_T-\Pi_C \le 0$ against $H_1: \Pi_T-\Pi_C>0$. Therefore, $\Delta_1=0$. We consider the following prior distribution for $\Pi_T - \Pi_C$  
\[
    \Pi_T - \Pi_C  \sim  {\rm Normal}\left[ \Delta_0=0.20, \;\; \sigma_0^2=0.06 \right]
\]
The allocation ratio is 2:1, resulting in $r^2=(2+1)^2/2=4.5$. At the final analysis, the SD and $\mbox{SE}(\delta_n)$ are projected as $\tilde{\sigma}=\sqrt{(2/3)\cdot (0.30*0.70/2+0.10*0.90)}=0.361$ and $\tilde{k}=\sqrt{4.5}(0.361)/\sqrt{210}=0.053$, respectively. Therefore, the PoS for trial success ($\gamma=c(1)=2.012$) and clinical success ($\gamma=\theta_{\min}/\tilde{k}=0.15/0.053=2.83$) at the design stage are $\Phi\left(\frac{\sqrt{210}(0.20-0)-\sqrt{4.5}(0.361)(2.012)}{\sqrt{(210)(0.06)+(4.5)(0.361)^2}}\right)=0.645$ and $\Phi\left(\frac{\sqrt{210}(0.20-0)-\sqrt{4.5}(0.361)(2.83)}{\sqrt{(210)(0.06)+(4.5)(0.361)^2}}\right)=0.578$, respectively (see Eq.~\eqref{eq:pp3cont}).  \\

Now consider the following interim results: 37.9\% of the patients responded in the luspatercept arm ($n_T=105$) compared to 22.2\% in the placebo arm ($n_C=53$). Therefore,  $\delta_n=0.379-0.222 = 0.157$   with SE as $\sqrt{0.379*0.621/105+0.222*0.778/53}=0.074$ and $s_n=SE\cdot \sqrt{n}/r=(0.074)(\sqrt{158})/\sqrt{4.5}=0.4385$. Further, $k=0.074\cdot\sqrt{0.75}=0.064$.Projecting the observed difference in proportion from the post-interim data as 0.20 (=$\Delta'$), the CP for trial success ($\gamma=c(1)=2.012$) and clinical success ($\gamma=0.15/0.064=2.34$) are 0.884 and 0.709, respectively (Eq.~\eqref{eq:cpcont}). Assuming the interim trend for remaining part of the trial, the CP for trial success and clinical success are 0.804 and 0.587, respectively (Eq.~\eqref{eq:cp2cont}).  The PPoS for trial success  and clinical success based on interim results and prior knowledge are 0.782 and 0.586, respectively (Eq.~\eqref{eq:ppcont}).  If we leave out the prior distribution, the PPoS for trial success  and clinical success are 0.772 and 0.575, respectively (Eq.~\eqref{eq:pp2cont}). \\

Plots of the predictive distributions of $\Delta=\Pi_T -\Pi_C$, and the CP and PPoS against $\delta_n$ (i.e., interim estimate of difference in proportion) are plotted presented in Figure~\ref{fig:example2}.  We can confirm that  CP$>$PPoS for  $CP>0.5$ and CP$<$PPoS for  $CP<0.5$ regardless the type of success (i.e., trial success or clinical success) when prior information are not incorporated in the PPoS.


\subsection{Survival endpoint, single-arm trial}\label{sec:survsingle}

\begin{figure}
\centering
\caption{Comparison of empirical SE of log(KM estimate of median) with the $1/d$ with $1/\log{2} \cdot 1/d$ as presented in Section~\ref{sec:survsingle}}
\label{fig:median_sd}
  \includegraphics [angle=0,width=120mm, height=80mm]{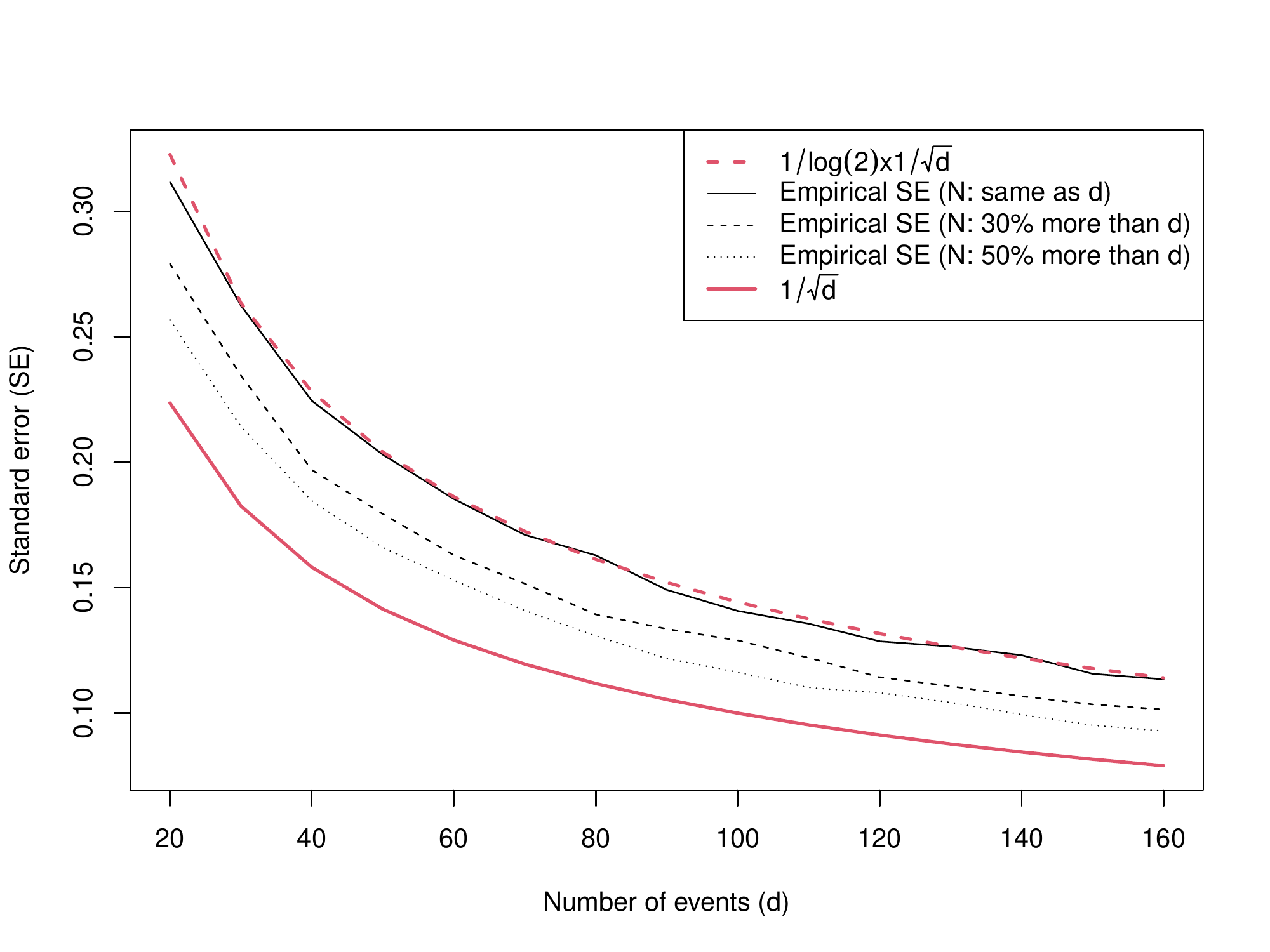}
\end{figure}

Let's consider a study with a single treatment arm and time-to-event endpoint.  Denote the population median as $M$ and the maximum number of events in the study as $D$.  We test the following hypotheses:
\[
H_0: M=M_1 \qquad {\rm vs.} \qquad H_1: M>M_1
\]
Here, $\theta =\log{M}-\log{M_1}$. At interim analysis with $d$ ($< D$) events, the estimate of $\theta$ is $\hat{\theta}(t)=\log{m_d}-\log{M_1}$ where $m_d$ is the estimated median at interim. Assuming that $m_d$  normally distributed (e.g., see \cite{brookmeyer1982confidence, owzar2008designing}), and with $\mbox{var}(\log{m_d})=\xi^2/d$, a test can by constructed as follows:
\[
Z(t)=\frac{\log{m_d}-\log{M_1}}{\xi/\sqrt{d}}=\frac{\log{(m_d/M_1)}\cdot\sqrt{d}}{\xi}
\]
Under the exponential time-to-event distribution, $\mbox{var}(\log{m_d})=1/d$ (i.e., $\xi=1$) if $m_d$ is maximum likelihood estimate (MLE) (see Appendix 1a), or $\mbox{var}(\log{m_d})=(\log{2})^{-2}/d$ (i.e., $\xi=(\log{2})^{-1}=1.443$) if $m_d$ is the plain sample median (see Appendix 1b). Further, $\mbox{var}(\log{m_d})=(\log{2})^{-2}\cdot \beta^{-2}/d$ (i.e., $\xi=(\log{2})^{-1}\cdot \beta^{-1}=1.443\cdot \beta^{-1}$) if $m_d$ is the plain sample median from a Weibull time-to-event distribution with $\beta$ shape parameter (see Appendix 1b). For other estimator (e.g., smallest time with at least 50\% Kaplan-Meier (KM) estimate of survival probability), $\xi^2$ is simply the ratio of the variance of the log of estimated median to that of MLE. We have plotted the empirical SE of log(KM estimate of median) for various event sizes (ranging between 20 and 60) and sample sizes (0\%, 30\% and 50\% more than event size) in Figure~\ref{fig:median_sd}. Empirical SE were obtained as follows: (a) 5000  datasets were simulated, (b) in each of the simulated datasets, event times were generated from exponential distribution with median of 12 months and KM estimates for median were obtained, and (c) SD of log(KM estimate of median) were obtained. The results suggest that $1/\sqrt{d}$ under-estimates the empirical SE. In comparison, $(\log{2})^{-1}/\sqrt{d}$ almost coincides with the  empirical SE for $N=d$. However, as $N$ becomes greater than $d$, $(\log{2})^{-1}/\sqrt{d}$ tends to over-estimate the SE as the follow-up times of the censored subjects attributes to the increased precision of KM estimates which was not factored into the derivation of $(\log{2})^{-1}/\sqrt{d}$.  In any case, $t=d/D$ and $k=\xi/\sqrt{D}$.\\ 

\textsl{Conditional power (CP)}: The CP with the projected median from  post-interim data as $M'$  is
\begin{equation} \label{eq:cpsurvsingle} 
       \Phi\left(\frac{1}{\xi}\sqrt{\frac{D}{D-d}}\left[\frac{1}{\sqrt{D}}\left\{d\cdot\log{\frac{m_d}{M_1}}+(D-d)\cdot\log{\frac{M'}{M_1}}\right\}-\xi \cdot \gamma\right]\right)
\end{equation}
Projecting the future trend to be identical with the interim trend (i.e., $M'=m_d$), the expression of the CP simplifies to
\begin{equation} \label{eq:cp2survsingle} 
       \Phi\left(\frac{1}{\xi}\sqrt{\frac{D}{D-d}}\left[\sqrt{D}\cdot\log{\frac{m_d}{M_1}}-\xi \cdot \gamma\right]\right)
\end{equation}

\textsl{Predictive power of success (PPoS)}: The PPoS solely based on the interim data can be expressed as
\begin{equation}\label{eq:ppsurvsingle}
       \Phi\left(\frac{1}{\xi}\sqrt{\frac{d}{D-d}}\left[\sqrt{D}\cdot\log{\frac{m_d}{M_1}}-\xi \cdot \gamma\right]\right) 
\end{equation}
Now, incorporating prior information specified in Eq.~\eqref{eq:priorsurvsingle}, expression of the PPoS is refined as
\begin{equation} \label{eq:pp2survsingle} 
    \Phi\left(
    \frac{1}{\xi}\sqrt{\frac{d}{D-d}}
    \frac
    {\left[(1-\psi)\{\frac{d}{\sqrt{D}}\cdot \log{\frac{m_d}{M_1}}
        +\frac{D-d}{\sqrt{D}}\cdot \log{\frac{M_0}{M_1}}\} + \psi \cdot \sqrt{D} \cdot \log{\frac{m_d}{M_1}} 
        - \xi \cdot \gamma \right]}
    {\sqrt{\psi+ (1-\psi)\frac{d}{D}}}\right)
\end{equation}
where, $\psi=d \cdot \sigma_0^2/(d \cdot \sigma_0^2+ \xi^2)$.\\

\textsl{Probability of success (PoS)}: The PoS of a prospective trial with $D$ events (therefore, $\tilde{k}=\xi/\sqrt{D}$) and the prior information specified in Eq.~\eqref{eq:priorsurvsingle} can be expressed as follows:
\begin{equation}
        \Phi \left( \frac{\sqrt{D} \cdot \log{(M_0/M_1)} - \xi \cdot \gamma}{ \sqrt{D\cdot \sigma_0^2 + \xi^2}} \right)   \label{eq:pp3survsingle} 
\end{equation}
For the calculation of the PPoS and PoS with prior distribution, the following prior for $M$ was used
\begin{equation}\label{eq:priorsurvsingle}
    \log{M}  \sim  {\rm Normal}\left[ \log{M_0}, \sigma_0^2 \right]
\end{equation}


\subsection{Survival endpoint, two-arm trial}\label{sec:surv}

Denoting the  treatment to control HR as $\Delta$, in a two-arm clinical trial with time-to-event endpoint and the maximum target number of events as $D$, we  test the following hypotheses:
\[
H_0: \Delta=\Delta_1 \qquad {\rm vs.} \qquad H_1: \Delta<\Delta_1
\]

Here, $\theta =\log{(\Delta_1/\Delta)}$. At interim analysis with the total $d$ events,  $\hat{\theta}(t)=\log{(\Delta_1/\delta_d)}$ where $\delta_d$ is the estimated HR.  The corresponding log-rank statistic for trend test is approximately equivalent to
\[
Z(t)=\frac{\log{(\Delta_1/\delta_d)}}{r}\sqrt{d}
\]
Further, in this case, $t=d/D$ and $k=r /\sqrt{D}$.\\ 

\textsl{Conditional power (CP)}: With the projected HR from  future data as $\Delta'$, the expression of CP is 
\begin{equation}\label{eq:cpsurv}
       \Phi\left(\frac{1}{r}\sqrt{\frac{D}{D-d}}\left[  \frac{d}{\sqrt{D}}\log{\frac{\Delta_1}{\delta_d}}
       +\frac{D-d}{\sqrt{D}}\log{\frac{\Delta_1}{\Delta'}}
       -r \cdot \gamma\right]\right)
\end{equation}
If the interim trend continues to hold in post-interim data (i.e., $\Delta'=\delta_d$),   the CP reduces to
\begin{equation}\label{eq:cp2surv}
       \Phi\left(\frac{1}{r}\sqrt{\frac{D}{D-d}}\left[  \sqrt{D} \cdot \log{\frac{\Delta_1}{\delta_d}}-r \cdot \gamma\right]\right)
\end{equation}

\textsl{Predictive power of success (PPoS)}: The PPoS solely based on the interim data can be expressed as 
\begin{equation}
           \Phi\left(\frac{1}{r}\sqrt{\frac{d}{D-d}}\left[  \sqrt{D} \cdot \log{\frac{\Delta_1}{\delta_d}}-r \cdot \gamma\right]\right) \label{eq:ppsurv}  
\end{equation}
Incorporating the prior information specified in Eq.~\eqref{eq:priorsurv}, revised expression of PPoS is \cite{tang2015optimal}
\begin{equation} \label{eq:pp2surv} 
    \Phi\left(
    \frac{1}{r}\sqrt{\frac{d}{D-d}}
    \frac
    {\left[(1-\psi)\{\frac{d}{\sqrt{D}}\cdot \log{\frac{\Delta_1}{\delta_d}}
        +\frac{D-d}{\sqrt{D}}\cdot \log{\frac{\Delta_1}{\Delta_0}}\} + \psi \cdot \sqrt{D} \cdot \log{\frac{\Delta_1}{\delta_d}} 
        - r \cdot \gamma \right]}
    {\sqrt{\psi+ (1-\psi)\frac{d}{D}}}\right)
\end{equation}
where, $\psi=\dfrac{d\cdot \sigma_0^2}{d\cdot \sigma_0^2 + r^2}$. \\

\textsl{Probability of success (PoS)}: The PoS of a prospective trial with $D$ events (hence, $\tilde{k}=r/\sqrt{D}$) and the prior information specified in Eq.~\eqref{eq:priorsurv} is (e.g., see \cite{wang2013evaluating}):
\begin{equation}
        \Phi \left( \frac{\sqrt{D} \cdot \log{(\Delta_1/\Delta_0)} - r \cdot \gamma}{ \sqrt{D\cdot \sigma_0^2 + r^2}} \right)   \label{eq:pp3surv} 
\end{equation}
For the calculation of the PPoS and PoS with prior distribution, the following prior for $\Delta$ was used
\begin{equation}\label{eq:priorsurv}
    \log{\Delta}  \sim  {\rm Normal}\left[ \log{\Delta_0}, \sigma_0^2 \right]
\end{equation}

\textbf{Example 3:} In the INTELLANCE-I trial on glioblastoma patients comparing investigational drug depatuxizumab mafodotin, total of 639 subjects (=$N$) were enrolled  with 1:1 allocation ratio \cite{lassman2020depatuxizumab}. The primary endpoint in the study was overall survival. The target number of events at the final analysis was 441 (=$D$) and an interim analysis was planned with 332 events. The trial used a weighted log-rank test, however, here we illustrate assuming standard log-rank test. Therefore, expected SE of $\log{HR}$ at final analysis is $k=2/\sqrt{441}=0.0952$. The  rejection boundaries for the Z test (i.e., trend test) statistic are 2.340 and 2.012 at interim and final analyses, respectively. For clinical success, we assume $\mbox{HR}\le 0.80$ (=$\Delta_{\min}$) resulting in  $\gamma=-\log{(0.80)}/0.0952=2.344$.  \\

\begin{figure}
\centering
\caption{Predictive distribution of HR (i.e., $\Delta$) and the relationship of CP and PPoS for trial and clinical success with $\delta_d$ (i.e., interim estimate of HR) based on the Example 3. Horizontal and vertical lines in the left panel correspond to 50\% power and observed HR of 0.82 (=$\delta_d$), respectively.}
\label{fig:example3}
\begin{subfigure}{.49\textwidth}
  \centering
  \includegraphics [angle=0,width=80mm, height=80mm]{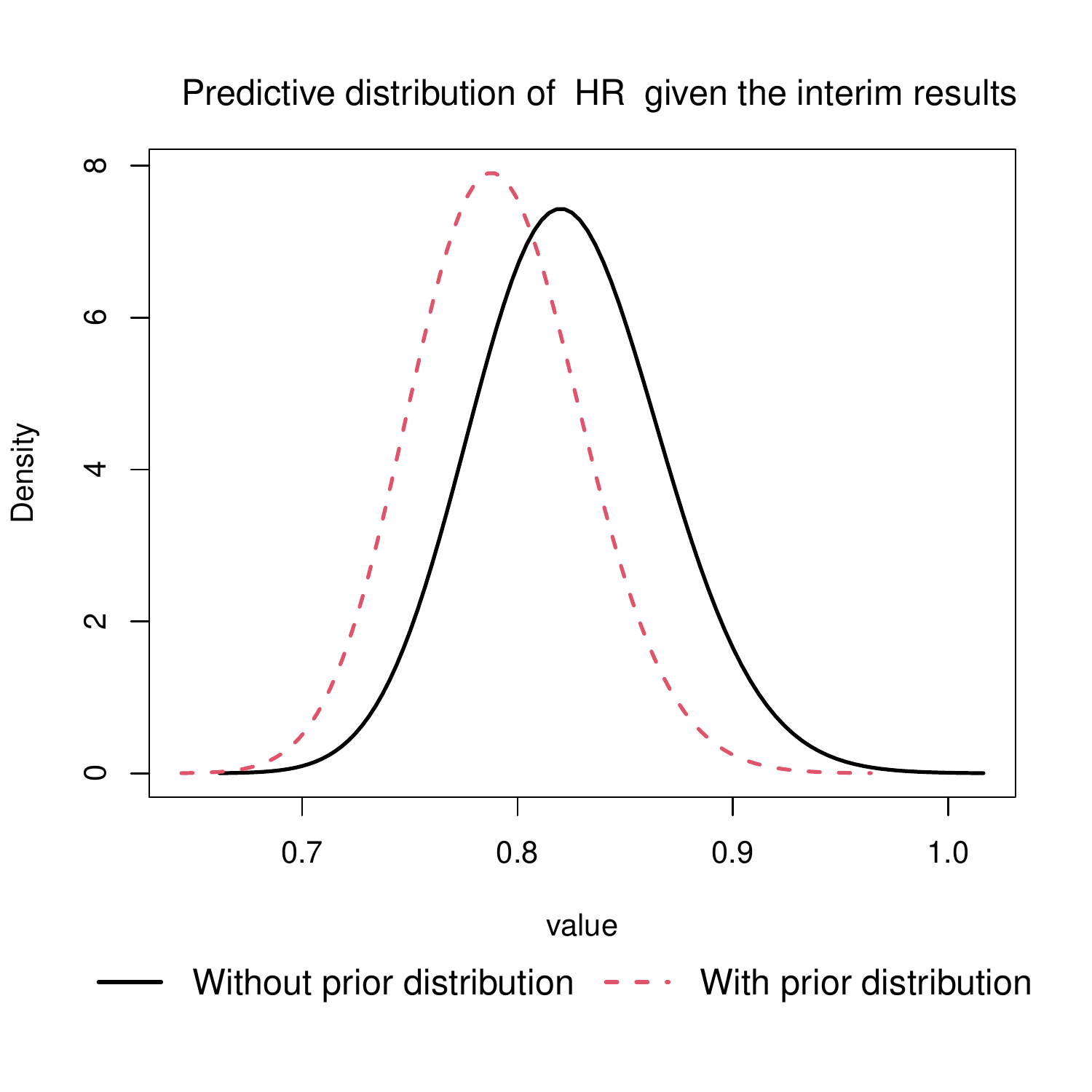}
\end{subfigure}
\begin{subfigure}{.49\textwidth}
  \centering
\includegraphics [angle=0,width=80mm, height=90mm]{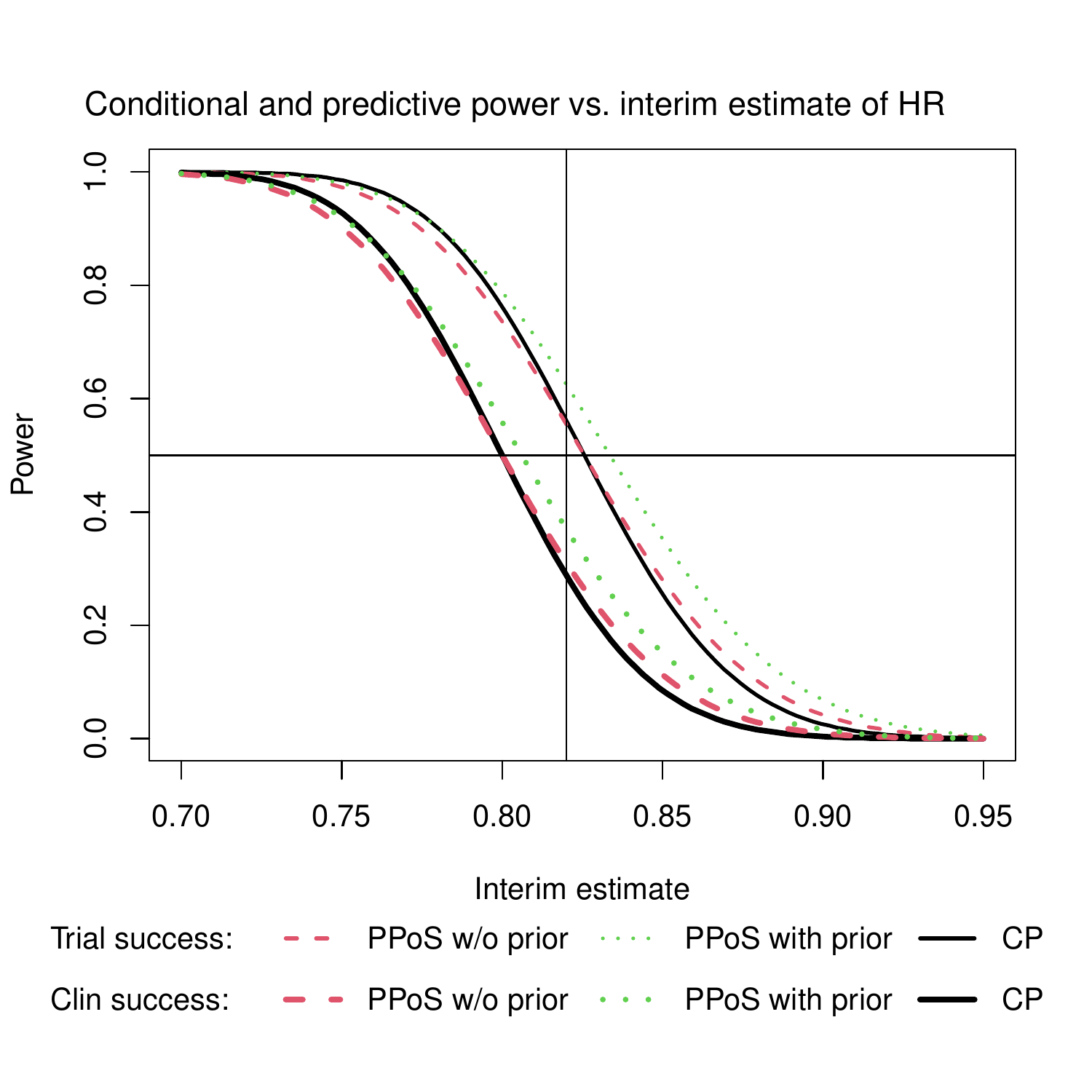}
\end{subfigure}
\end{figure}

We test following hypotheses: $H_0: {\rm HR}=1$ vs. $H_1: {\rm HR}<1$  (i.e., $\Delta_1=1$). The phase 2 trial on recurrent disease \cite{van2020intellance} reported HR of 0.71 (=$\Delta_0$) with 133 events  resulting in prior distribution 
\[
\log{\Delta}  \sim  {\rm Normal}\left[ \log{0.71}, \sigma_0^2=(2/\sqrt{133})^2=(0.173)^2 \right]
\]
Due to 1:1 allocation ratio, $r^2=(1+1)^2/2=4$. Thus, the PoS for trial success ($\gamma=1.96$) and clinical success ($\gamma=2.344$) are  $\Phi\left(\frac{\sqrt{441}\log{(1/0.71)}-(2)(1.96)}{\sqrt{(441)(0.173)^2+(2)^2}}\right)=0.785$  and  $\Phi\left(\frac{\sqrt{441}\log{(1/0.71)}-(2)(2.344)}{\sqrt{(441)(0.173)^2+(2)^2}}\right)=0.727$, respectively, at the design stage of a single-look trial without interim analysis (Eq.~\eqref{eq:pp3surv}).\\

Consider the following interim results: estimated HR of 0.82 (=$\delta_d$) with 346 events (=$d$).  Note that, $k=2/\sqrt{441}=0.0952$.  Assuming the projected HR from the post-interim data as 0.75 (=$\Delta'$) as assumed at the design stage, the CP for trial success ($\gamma=2.012$) and clinical success ($\gamma=2.344$) are 0.722 and 0.451, respectively (see Eq.~\eqref{eq:cpsurv}). However, if we assume that the interim trend continues to the remaining part of the trial, the CP for trial success and clinical success are 0.561 and 0.288, respectively (see Eq.~\eqref{eq:cp2surv}).  The PPoS for trial success  and clinical success solely based on the interim results are 0.554 and 0.310, respectively (see Eq.~\eqref{eq:ppsurv}). If we incorporate prior distribution, the PPoS for trial success  and clinical success are 0.625 and 0.370, respectively (see Eq.~\eqref{eq:pp2surv}).  \\

In Figure~\ref{fig:example3}, the predictive distribution of HR are plotted in the left panel, and the CP and PPoS against the interim HR estimate are plotted in the right panel. For both trial success and clinical success,  CP$>$PPoS for  $CP>0.5$ and CP$<$PPoS for  $CP<0.5$ when prior information is ignored.

\section{Binary endpoint with beta prior}\label{sec:betabinom}

So far we have discussed normally distributed test statistics and normal priors. Here, we  discuss the derivation of PPoS in trials with binary endpoint with beta prior.  We have considered the same notations and hypothesis testings presented in Section~\ref{sec:binarysingle} and Section~\ref{sec:binarytwo}. 

\subsection{single-arm trial}

Consider a prior distribution of probability of response $\Pi$ as ${\rm Beta}(a, b)$. Denoting the observed number of response as $x_n$ from $n$ subjects at interim analyses,  the posterior distribution of $\Pi$  is 
\begin{equation*}
    \Pi\;|\; x_n \sim {\rm Beta}(x_n + a, n - x_n + b)
\end{equation*}
 Let, $Y$ be the number of the observed response from remaining $N-n$ subjects.  The predictive distribution of $Y$ is (e.g., see \cite{choi1985early, johns1999use})
 \begin{equation*}
     \mbox{Pr}(Y=y|x_n) = {N-n \choose y} \frac{{\rm B}(x_n+y+a,\; N-x_n-y+b)}{{\rm B}(x_n+a, \;n-x_n+b)} \qquad  y=0, 1, \cdots, N-n
 \end{equation*}
where, ${\rm B}(u, v)=\dfrac{(u-1)!(v-1)!}{(u+v-1)!}$ is the beta function. Thus, the PPoS would be 
\begin{equation}
   \sum_{y=0}^{N-n}{I({\rm success}|x_n+y, N)\cdot {\rm Pr}(Y=y|x_n)}
\end{equation}
where $I(\cdot)$ is the indicator function for trial success (e.g., based on approximate Z test or exact binomial test) or  clinical success (i.e., estimated proportion exceeds  certain threshold value) is met.

\subsection{two-arm trial}

We assume following priors for the proportions in treatment (T) and control (C) arms: $\Pi_T \sim {\rm Beta}(a_T, b_T)$ and $\Pi_C \sim {\rm Beta}(a_C, b_C)$. At the interim analysis, $x_T$   of $n_T$ subjects in the treatment arm and $x_C$  of  $n_C$ subjects  in the control arm are responded. The posterior distributions are  \cite{saville2014utility}
\begin{equation*}
    \Pi_T\;|\; x_T \sim {\rm Beta}(x_T + a_T, n_T - x_T + b_T)
\end{equation*}
\begin{equation*}
    \Pi_C\;|\; x_C \sim {\rm Beta}(x_C + a_C, n_C - x_C + b_C)
\end{equation*}

 Let, $Y_T$ and $Y_C$ be the number of responders from remaining $N_T-n_T$ and $N_C-n_C$ subjects, respectively.  The predictive distribution of $Y_T$ and $Y_C$ are (e.g., see \cite{johns1999use})
 \begin{equation*}
     {\rm Pr}(Y_T=y_T|x_T) = {N_T-n_T \choose y_T} \frac{{\rm B}(x_T+y_T+a_T,\; N_T-x_T-y_T+b_T)}{{\rm B}(x_T+a_T, \;n_T-x_T+b_T)} \qquad  y_T=0, 1, \cdots, N_T-n_T
 \end{equation*}
 \begin{equation*}
     {\rm Pr}(Y_C=y_C|x_C) = {N_C-n_C \choose y_C} \frac{{\rm B}(x_C+y_C+a_C,\; N_C-x_C-y_C+b_C)}{{\rm B}(x_C+a_C, \;n_C-x_C+b_C)} \qquad  y_C=0, 1, \cdots, N_C-n_C
 \end{equation*}
Thus, the PPoS  at the end of the trial  is 
\begin{equation} \label{eq:pp2betabinom} 
    \sum_{y_T=0}^{N_T-n_T}
                 \sum_{y_C=0}^{N_C-n_C}{
                 I({\rm success}|x_T+y_T, x_C+y_C, N_T, N_C)\cdot
                {\rm Pr}(Y_T=y_T|x_T) \cdot
                {\rm Pr}(Y_C=y_C|x_C)}
\end{equation}
where $I(\cdot)$ is the indicator function for success criteria which could be either trial success (e.g., based on approximate Z test or Fisher's exact test) or  clinical success indicating the observed difference in proportion exceeds the certain clinically meaningful value.\\

\textbf{Example 4:} This example is inspired by the example given in \citet{johns1999use}. Consider a clinical trial to demonstrate that the relapse rate in patients treated in the experimental treatment arm is less than the control arm's response rate. It was planned to enrol 340 patients in each arm. The interim analysis was planned after 170 patients in each arm completed treatment. Non-informative uniform priors were assumed for the two relapse rate: $\Pi_T \sim {\rm Beta}(a_T=1, b_T=1)$ and $\Pi_C \sim {\rm Beta}(a_C=1, b_C=1)$.\\

In this case, we are statistically testing the following hypotheses: $H_0: \Pi_T-\Pi_C \le 0$ against $H_1: \Pi_T-\Pi_C>0$.  Suppose we observed following results at interim analysis: (a) in the treatment arm, 155 (=$n_T$) out of 170 patients responded with 13 (=$x_T$) subsequent relapses, and (b) in the control arm, 152 (=$n_C$) out of 169 patients responded with 21 (=$x_C$) subsequent relapses. Subsequently, additional $340-170=170$ patients (=$N_T-n_T$) and $340-169=171$ patients (=$N_C-n_C$) to be enrolled in the treatment arm and control arm, respectively. With this information, the PPoS for trial success based on a Z test at one sided 0.025 level is 0.536 (see Eq.~\eqref{eq:pp2betabinom}).

\section{Software implementation}\label{sec:software}

\begin{figure}
\centerline{%
\includegraphics [angle=0,width=180mm, height=200mm]{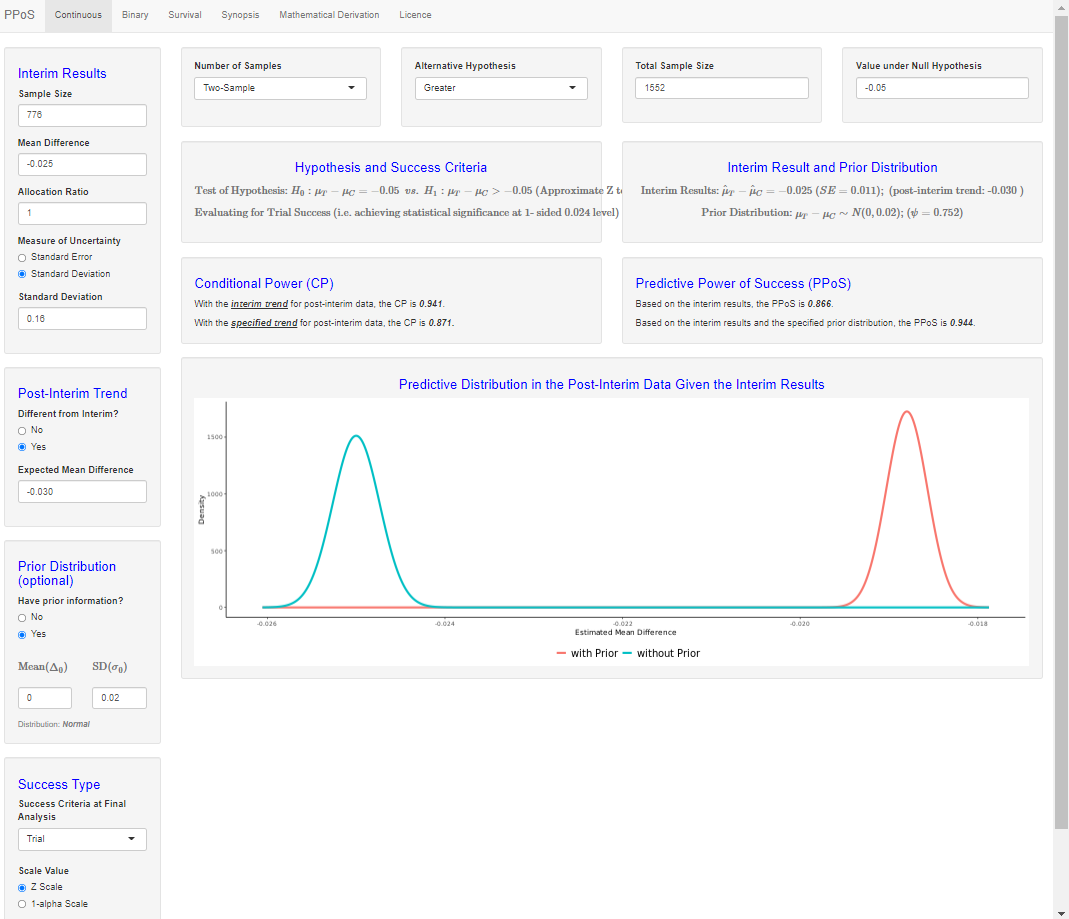}}
\caption{R shiny app for calculation of CP, PPoS and PoS}
\label{fig:shiny}
\end{figure}

The expressions of CP, PPoS and PoS presented in this paper are implemented in \verb|LongCART| package in \verb|R|: (a) \verb|PoS()| to calculate PoS at the design stage, (b) \verb|succ_ia()| to calculate CP and PPoS, and (c) \verb|succ_ia_betabinom_one()| and \verb|succ_ia_betabinom_two()|  to calculate PPoS for binary endpoint with beta prior. Examples of these functions are provided in the Appendix 2. A user-friendly R shiny app is also available at https://ppos.herokuapp.com/ to calculate these measures. A screenshot of this shiny app is presented in Figure~\ref{fig:shiny}.

\section{Discussion}

In this paper, expressions for various measures of the probability of success are presented by type of endpoints. The discussion in this paper is restricted to the normally distributed test statistics along with normal prior, and beta prior for binomial distributions. For other distributions, the relevant expressions can be obtained using the general framework presented in Section~\ref{sec:prelim} or simulation based methods such as Bayesian clinical trial simulation (BCTS) \cite{o2005assurance, wang2013evaluating} may be used. Nevertheless, a natural question arises which  pr(success) measure one should prefer. Often PPoS is preferred over CP for the following reasons: (1) these have better predictive interpretation, (2) unlike frequentist counterpart, the knowledge on $\theta$ (the parameter of interest) is used as distribution, whereas in frequentist calculation, we assume that the value of $\theta$ is known without any uncertainty, and (3) unlike frequentist paradigm, the prior information can be incorporated in the Bayesian paradigm.  As evident from Figure~\ref{fig:example1}, Figure~\ref{fig:example2} and Figure~\ref{fig:example3}, the CP is more aggressive than the PPoS and hence use of the CP  increases the chance of early stopping for futility or efficacy. \citet{lachin2005review} has shown that futility termination may markedly decrease the power in direct proportion to the probability of stopping for futility. Therefore, PPoS seems to be more useful while monitoring a trial for early termination. \\

\citet{low2011perils} discussed the disadvantages of PoS and  PPoS, especially PPoS can lead to much larger sample sizes than the CP during sample size re-estimation. On the other hand, the effect of varying prior distribution of $\theta$ on predictive power in the context of futility monitoring is discussed by \citet{dmitrienko2006bayesian}, and in general by \citet{rufibach2016bayesian}. In summary, they have proposed to use aggressive prior for futility monitoring as the use of non-informative may increase the early termination rate. \citet{tang2015optimal} suggested using the upper limit of PPoS in futility monitoring. The effect of prior on PPoS in the context of the binomial endpoint is discussed by \citet{johns1999use}.\\

One might consider the predictive power of clinical success (PPoCS) in monitoring for early stopping as well; however,  in general, its use should be discouraged.   \citet{saville2014utility}  have pointed out that the PPoS with respect to trial success (referred to as 'predictive probabilities') are naturally appealing for monitoring a clinical trial as (a) the PPoS directly addresses the question of whether the study is going to be a success at the end, (b) and the PPoS often changes drastically with the accrual of more data whereas the PPoCS (referred as 'posterior probabilities') may remain nearly identical. Further, we also would like to point out the potential misuse of PPoS for survival endpoints with delayed treatment effects. In that case, the use of futility criteria for early stopping based on PPoS or CP may be misleading. In these cases, the futility criteria, if any, must be determined through exhaustive evaluation of operating characteristics.

\setcounter{secnumdepth}{0}

\section{Appendix}

\subsection{Appendix 1: Variance of median estimates in a single-arm trial with time-to-event endpoint}

\subsubsection{Appendix 1a: Variance of MLE of median with exponential time-to-event distribution}

For an exponential time-to-event distribution with density $f(x)=\lambda \exp{(-\lambda x)}$ and $d$ events, the asymptotic variance of $\hat{\lambda}$ (MLE of $\lambda$) is estimated by $\hat{\lambda}^2/d$, so $1/d$ estimates the variance of $\log{\hat{\lambda}}$  \cite{brookmeyer1982confidence}. The MLE of median can be expressed as $m_d=\log{2}/\hat{\lambda}$. Therefore, the variance of $\log{m_d}$ is
\[
\mbox{var}[\log{m_d}]=\mbox{var}[\log{(\log{2})}-\log{\hat{\lambda}}]=\mbox{var}[\log{\hat{\lambda}}]=1/d
\]

\subsubsection{Appendix 1b: Variance of observed sample median with Weibull and exponential time-to-event distribution}

Consider a Weibull time-to-event distribution with density $f(x)=\beta \lambda x^{\beta-1}\exp{(-\lambda x^{\beta-1})}$. The corresponding cumulative distribution function (CDF) is $F(x)=1-\exp{(-\lambda x^{\beta-1})}$. Equating this CDF with 0.5, we obtain the median value as $M=(\log{2}/\lambda)^{1/\beta}$. The density at $M$ is 
\[
f(M)=\frac{1}{2}\cdot \beta \cdot \lambda \cdot \left(\frac{\log{2}}{\lambda}\right)^{\frac{\beta-1}{\beta}}
\]
If $m_d$ is the sample median based on $d$ events (and, entirely ignoring the censored subjects), estimated variance of $m_d$ is \cite{Miller2016supplement}
\[
\mbox{var}(m_d)= \frac{1}{4\cdot d \cdot f(M)}=\frac{1}{d} \cdot \frac{1}{\beta^{2}} \cdot \frac{1}{\lambda^{2}} \cdot 
\left(\frac{\lambda}{\log{2}}\right)^{2\cdot \frac{\beta-1}{\beta}}
\]
Consequently, the variance of $\log{m_d}$ is
\[
\mbox{var}[\log{m_d}]= \left(\frac{1}{m_d}\right)^2\mbox{var}(m_d)=\frac{1}{d} \cdot \frac{1}{\beta^{2}}  \cdot 
\left(\frac{1}{\log{2}}\right)^2
\]
For $\beta=1$ (i.e., exponential underlying distribution), $\mbox{var}[\log{m_d}]=d^{-1}\cdot (\log{2})^{-2}$.

\subsubsection{Appendix 1c: R codes used to compare empirical SD of log(KM estimate of median) with the $1/d$ with $1/\log{2} \cdot 1/d$ in Figure~\ref{fig:median_sd}}
\begin{small}
\begin{lstlisting}[language=R]
library(survival)

sim.fn<- function(N, D, med, M, ltfu.rate=0.05){
lambda.event<- log(2)/med 
lambda.censor<- 1/(1/ltfu.rate - 1)*lambda.event

event.t<- rexp(n = N*M, rate = lambda.event)
censor.t<- rexp(n = N*M, rate = lambda.censor)
replicate<- kronecker(1:M, rep(1,N))
dat<- data.frame(replicate, event.t, censor.t)
dat<- transform(dat, fup=pmin(event.t, censor.t), event=I(event.t<=censor.t)*1)
dat.event<- subset(dat, event==1)
dat.event.sort<- dat.event[order(dat.event$replicate, dat.event$fup),] 
myfn<- function(x, d) x[min(length(x),d)]  #myfn(1:20, 10)
max.event.t<- tapply(dat.event.sort$fup, dat.event$replicate, FUN=myfn, d=D)
dat$max.event.t<- kronecker(max.event.t, rep(1,N))
dat<- transform(dat, fup=pmin(fup, max.event.t), event2=I(fup<=max.event.t)*event)

fit <- survfit(Surv(fup, event2) ~ strata(replicate), data = dat)
medians.km<- summary(fit)$table[,"median"]
names(medians.km)<- NULL
sd.obs.median<- sd(log(medians.km)) 
sd.theory1<- 1/sqrt(D)
sd.theory2<- 1/(log(2))*1/sqrt(D)
ret<- c(N, D, med, sd.obs.median, sd.theory1, sd.theory2, ltfu.rate, M)
ret
}

#--- Example: sim.fn(N=round(80*1.30), D=80, med=12, M=5000, ltfu.rate=0.000005)
\end{lstlisting}
\end{small}

\subsection{Appendix 2: Example R codes}

\subsubsection{Appendix 2a: Calculation of PoS with normal approximation}
\begin{small}
\begin{lstlisting}[language=R]
#--- Example 1 in the paper (continuous endpoint)
PoS(type="cont", nsamples=2, null.value=-0.05, alternative="greater", 
        N=1552, a=1,  
        succ.crit="trial", Z.crit.final=1.97,
        se.exp=0.12*sqrt(1/776 + 1/776),
        meandiff.prior=0, sd.prior=0.02) 

#--- Example 3 in the paper (survival endpoint)
PoS(type="surv", nsamples=2, null.value=1, alternative="less", 
        D=441,  
        succ.crit="trial", Z.crit.final=1.96,
        hr.prior=0.71, D.prior=133) 

PoS(type="surv", nsamples=2, null.value=1, alternative="less", 
        D=441,  
        succ.crit="clinical", clin.succ.threshold =0.8,
        hr.prior=0.71, D.prior=133) 
\end{lstlisting}
\end{small}

\subsubsection{Appendix 2b: Calculation of CP and PPoS with normal approximation}
\begin{small}
\begin{lstlisting}[language=R]
#--- Example 1 in the paper (continuous endpoint)
succ_ia(type="cont", nsamples=2, null.value=-0.05, alternative="greater",
        N=1552, n=776, a=1,   
        meandiff.ia=-0.025, sd.ia=0.16,      
        succ.crit="trial", Z.crit.final=1.97,  
        meandiff.exp=-0.030, 
        meandiff.prior=0, sd.prior=0.02) 

#--- Example 2 in the paper (binary endpoint)
p1<- 0.379; p2<- 0.222
n1<- 105; n2<- 53

#-- Trial success
succ_ia(type="bin", nsamples=2, null.value=0, alternative="greater",
        N=210, n=158,  a=2,
        propdiff.ia=p1-p2,
        stderr.ia=sqrt(p1*(1-p1)/n1 + p2*(1-p2)/n2), 
        succ.crit="trial", Z.crit.final=2.012,
        propdiff.exp=0.20,
        propdiff.prior=0.20, sd.prior=sqrt(0.06))  

#-- Clinical success
succ_ia(type="bin", nsamples=2, null.value=0, alternative="greater",
        N=210, n=158,  a=2,
        propdiff.ia=p1-p2,
        stderr.ia=sqrt(p1*(1-p1)/n1 + p2*(1-p2)/n2), 
        succ.crit="clinical", clin.succ.threshold=0.15,
        propdiff.exp=0.20,
        propdiff.prior=0.20, sd.prior=sqrt(0.06)) 


#--- Example 3 in the paper (survival endpoint)

#--- Trial success
succ_ia(type="surv", nsamples=2, null.value=1, alternative="less", 
        D=441, d=346, a=1,   
        hr.ia=0.82,        
        succ.crit="trial", Z.crit.final=2.012,            
        hr.exp=0.75,
        hr.prior=0.71, D.prior=133) 

#--- clinical success
succ_ia(type="surv", nsamples=2, null.value=1, alternative="less", 
        D=441, d=346, a=1,   
        hr.ia=0.82,        
        succ.crit="clinical", clin.succ.threshold=0.80,            
        hr.exp=0.75,
        hr.prior=0.71, D.prior=133) 
\end{lstlisting}

\subsubsection{Appendix 2c: Calculation of PPoS for binary endpoint with beta-binomial approximation}

\begin{lstlisting}[language=R]
#--- Trial success
succ_ia_betabinom_two( N.trt=155+170, N.con=152+171,  
        n.trt=155, x.trt=13, n.con=152, x.con=21, 
        alternative="less", test="z",
        succ.crit = "trial", Z.crit.final = 1.96,  
        a.trt = 1, b.trt=1, a.con=1, b.con=1) 

#--- clinical success
succ_ia_betabinom_two( N.trt=155+170, N.con=152+171,  
        n.trt=155, x.trt=13, n.con=152, x.con=21, 
        alternative="less", test="fisher",
        succ.crit = "trial", Z.crit.final = 1.96,  
        a.trt = 1, b.trt=1, a.con=1, b.con=1) 
\end{lstlisting}
\end{small}
\end{document}